\documentclass[12pt,preprint]{aastex}
\usepackage{epsfig}
\usepackage{natbib}
\usepackage{graphicx}
\usepackage{slashbox}
\usepackage{multirow}
\usepackage{lscape}
\usepackage{mathrsfs,amssymb}
\usepackage{amsmath}
\usepackage{subfigure}
\usepackage{amssymb}
\usepackage{multirow}
\usepackage{tabularx}
\usepackage{rotating}
\usepackage{amsmath}
\usepackage{ulem}

\newcommand       \Angstrom     {\,{\rm \AA}}

\newcommand       \cm           {\,{\rm cm}}

\newcommand       \erg          {\,{\rm erg}}

\newcommand       \K            {\,{\rm K}}

\newcommand       \s            {\,{\rm s}}

\newcommand       \simgt        {\gtrsim}

\newcommand       \mum          {\,{\rm \mu m}}

\newcommand       \Teff         {T_{\rm eff}}

\newcommand       \simali       {\sim\,}

\newcommand       \Aaro          {A_{3.3}}

\newcommand       \Aali           {A_{3.4}}

\newcommand       \Iratio         {I_{3.4}/I_{3.3}}
\newcommand       \Adfa       {A_{6.85}}

\newcommand       \Adfb       {A_{7.25}}

\newcommand       \Acc        {A_{6.2}}

\newcommand       \NC         {N_{\rm C}}
\newcommand       \NCaro      {N_{\rm C,aro}}
\newcommand       \NCali      {N_{\rm C,ali}}
\newcommand       \NHaro      {N_{\rm H,aro}}
\newcommand       \NHali      {N_{\rm H,ali}}
\newcommand       \alifrac      {\eta_{\rm ali}}

\newcommand       \CabsPAH    {C^{\scriptscriptstyle\rm PAH}_{\rm abs}}
\newcommand       \Iratiomod   {\left(I_{3.4}/I_{3.3}\right)_{\rm mod}}
\newcommand       \Iratioobs   {\left(I_{3.4}/I_{3.3}\right)_{\rm obs}}

\countdef\decade=200
\advance\decade by \year
\countdef\hours=201
\advance\hours by \time
\divide\hours by 60
\countdef\mins=202
\advance\mins by \hours
\multiply\mins by 60
\multiply\hours by 100
\countdef\miltime=203
\advance\miltime by \hours
\advance\miltime by \time
\advance\miltime by -\mins
\def\today{\number\decade.\number\month.\number\day.\number\miltime}

\shorttitle{Aliphatics and Aromatics in the Universe}
\title{
Aliphatics and Aromatics in the Universe: The Pre-JWST Era
\\{\small DRAFT: \today ~~}
}
\author{X.J.~Yang\altaffilmark{1,2} and
            Aigen Li\altaffilmark{2}
            }
\altaffiltext{1}{Department of Physics,
                      Xiangtan University,
                      411105 Xiangtan, Hunan Province, China;
                      {\sf xjyang@xtu.edu.cn}}
\altaffiltext{2} {Department of Physics and Astronomy,
                  University of Missouri,
                  Columbia, MO 65211, USA;
                  {\sf lia@missouri.edu}}

\begin{document}

\begin{abstract}
The so-called ``unidentified infrared emission'' (UIE) features
at 3.3, 6.2, 7.7, 8.6, and 11.3$\mum$
ubiquitously seen in a wide variety of astrophysical regions
are generally attributed to polycyclic aromatic hydrocarbon
(PAH) molecules. Astronomical PAHs often have an {\it aliphatic}
component (e.g., aliphatic sidegroups like methyl --CH$_3$
may be attached as functional groups to PAHs)
as revealed by the detection in many UIE sources of
the {\it aliphatic} C--H stretching feature at 3.4$\mum$.
With its unprecedented sensitivity,
unprecedented spatial resolution
and high spectral resolution,
the {\it James Webb Space Telescope} (JWST)
holds great promise for revolutionizing
the studies of aliphatics and aromatics
in the universe. To facilitate analyzing
JWST observations, we  present a theoretical
framework for determining the aliphatic fractions
($\alifrac$) of PAHs, the fractions of C atoms
in aliphatic units, from the emission intensity ratios
of the 3.4$\mum$ aliphatic C--H feature
to the 3.3$\mum$ aromatic C--H feature.
To demonstrate the effectiveness of
this framework, we compile
the 3.3 and 3.4$\mum$ UIE data
obtained in the pre-JWST era
for an as complete as possible sample,
and then apply the framework to these pre-JWST data.
We derive a median aliphatic fraction of
$\langle\alifrac\rangle\approx 5.4\%$,
and find that the aliphatic fractions
are the highest in protoplanetary nebulae
illuminated by cool stars lacking ultraviolet radiation.
Nevertheless, the ``hardness'' of stellar photons
is not the only factor affecting the PAH aliphaticity,
other factors such as the starlight intensity
may also play an important role.
\end{abstract}


\section{Introduction\label{sec:intro}}
Polycyclic aromatic hydrocarbon (PAH) molecules,
composed of fused benzene rings,
have long been thought to be ubiquitous
in the interstellar medium (ISM),
as evidenced by a series of emission bands
observed at wavelengths 3.3, 6.2, 7.7, 8.6
and 11.3$\mum$, which are coincident with
the vibrational transitions of PAHs
(L\'eger \& Puget 1984, Allamandola et al.\ 1985).
These emission bands are often also known as
the ``unidentified infrared (IR) emission'' (UIE) bands.
Of all interstellar carbon, $\simali$15\%
is thought to be incorporated
into PAHs (Li \& Draine 2001).
Their emission accounts for up to 20\%
of the total IR power of the Milky Way
and star-forming galaxies (see Li 2020).

It has been generally held that astronomical
PAHs are not really  {\it pure} aromatic
compounds (see Kwok 2022).
They may include ring {\it defects},
{\it substituents}, partial {\it dehydrogenation}
and sometimes {\it superhydrogenation}
or {\it deuteration} (see Yang et al.\ 2017a
and references therein).
Astronomical PAHs often also include
an aliphatic component
(e.g., aliphatic sidegroups like methyl --CH$_3$
may be attached as functional groups to PAHs),
as revealed by the detection in many UIE sources
of a weak satellite emission feature
at 3.4$\mum$ which always accompanies
the 3.3$\mum$ emission feature
(see Yang et al.\ 2017b and references therein).
While the 3.3$\mum$ feature arises
from {\it aromatic} C--H stretch,
the 3.4$\mum$ feature is generally thought
to arise from {\it aliphatic} C--H stretch,
although it could also be due to
anharmonicity (Barker et al.\ 1987)
and superhydrogenation
(Bernstein et al.\ 1996, Yang et al.\ 2020).
In addition, some UIE sources also exhibit
two aliphatic C--H deformation bands
at 6.85 and 7.25$\mum$
(see Yang et al.\ 2016a and references therein).
Typically, for those sources with prominent
6.85 and 7.25$\mum$ bands,
the 3.4$\mum$ band is often pronounced.

Let $\alifrac\equiv\NCali/\left(\NCaro+\NCali\right)$
be the aliphatic fraction of PAHs, i.e., the ratio of
the number of C atoms in aliphatic units
($N_{\rm C,ali}$) to that in aromatic rings
($N_{\rm C,aro}$) {\it plus} that in aliphatic units.
In recent years, the PAH aliphatic fraction
has received increasing attention
(e.g., see Kwok \& Zhang 2011;
Li \& Draine 2012; Rouill\'e et al.\ 2012;
Steglich et al.\ 2013;
Pilleri et al.\ 2015;
Bernstein et al.\ 2017;
Buragohain et al.\ 2015, 2016, 2020;
Allamandola et al.\ 2021;
Yang et al.\ 2013, 2016a,b, 2017a,b).
%
Despite the widespread acceptance and extreme popularity
of the PAH model, the exact nature of the UIE carriers remains
unknown and many candidate materials have been proposed.
All these hypotheses generally agree that the UIE bands
arise from some sort of {\it aromatic} hydrocarbon material.
The major debate lies in the exact structure of the UIE carriers:
are they
(i) free-flying, predominantly {\it aromatic} gas-phase
    molecules like PAHs, or
(ii) amorphous solids
     (either bulk or nano-sized) with
     a {\it mixed aromatic/aliphatic} structure
     (e.g., see  Sakata et al.\ 1987,
     Papoular et al.\ 1993, Kwok \& Zhang 2011,
     Jones et al.\ 2013)?
One way to address this
is to examine the {\it aliphatic fraction} of the UIE carriers:
while PAHs, by definition, are predominantly {\it aromatic},
all other (proposed) carriers are considerably {\it aliphatic}
(see Yang et al.\ 2017b).

\begin{figure*}
\centering{
\includegraphics[scale=0.5,clip]{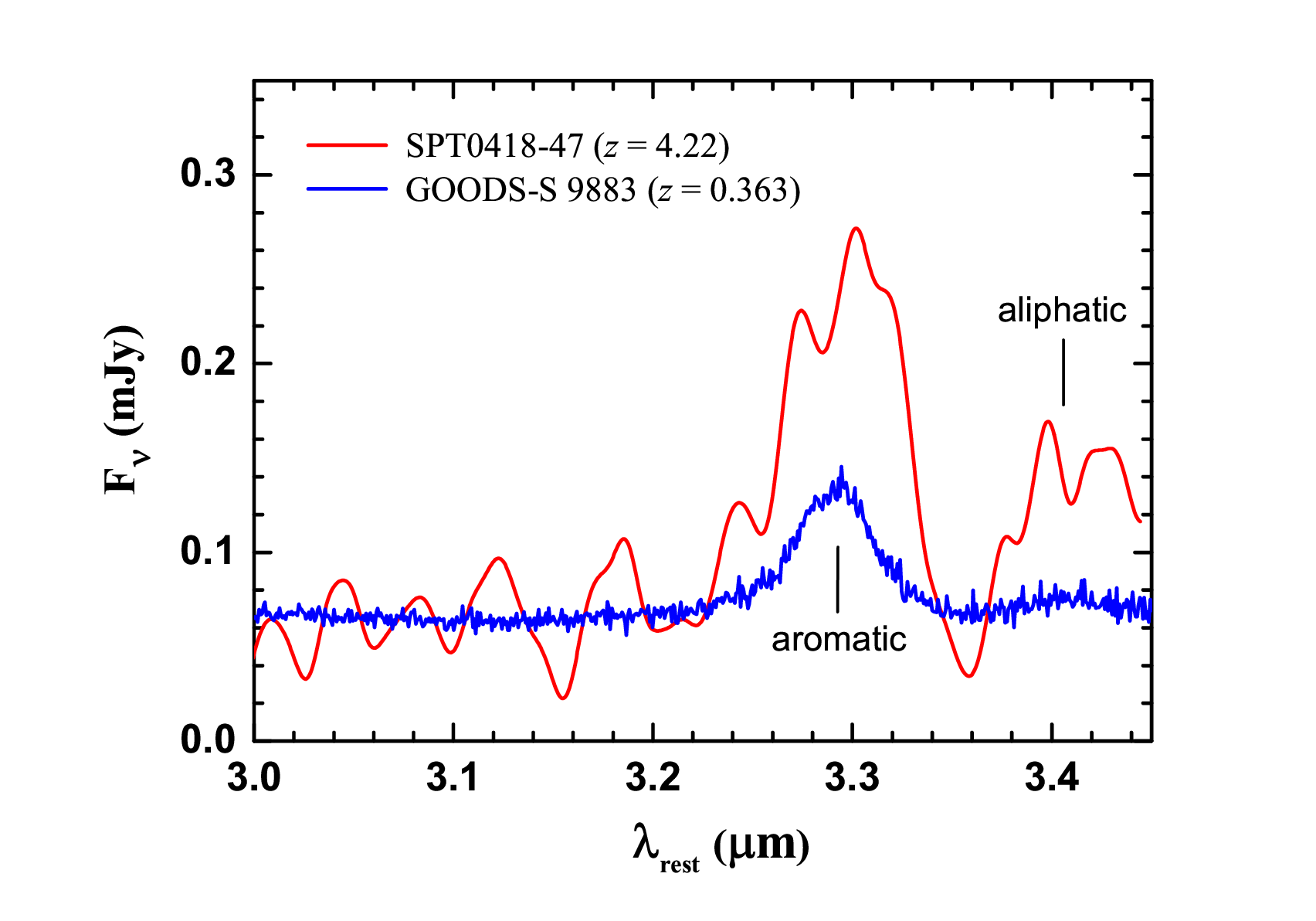}
}
\caption{\footnotesize
         \label{fig:jwst}
         The aromatic and aliphatic C--H stretches
         respectively at $\simali$3.3 and 3.4$\mum$
         from SPT0418-47, a galaxy at $z\approx4.22$,
         detected by JWST/MIRI (Spilker et al.\ 2023).
         The MIRI spectrum has been smoothed to
         a resolution of $R\approx600$.
         Also shown is the JWST/NIRCam spectrum
         of GOODS-S 9883, a moderately distant galaxy
         at $z$\,$\sim$\,0.36 (Lyu et al.\ 2023).
         To facilitate comparison, the JWST/NIRCam spectrum
         has been multiplied by a factor of three.
         }
\end{figure*}

Prior to the launch of
the {\it James Webb Space Telescope}
(JWST), the 3.4$\mum$ feature,
together with the 3.3$\mum$ feature,
has already been seen in a wide variety
of Galactic and extragalactic regions,
including reflection nebulae, H{\sc ii} regions,
photodissociated regions (PDRs),
protoplanetary nebulae, planetary nebulae,
protoplanetary disks around Herbig Ae/Be stars
and T Tauri stars, and external galaxies
(see Yang et al.\ 2017b).
Undoubtedly, the high spectral resolution
and unprecedented sensitivity of JWST will
bring the studies on aliphatics and aromatics
to a new height. Indeed, as illustrated in
Figure~\ref{fig:jwst}, the 3.3 and (tentatively) 3.4$\mum$
features were very recently seen in the mid-IR spectrum
of SPT0418-47, a galaxy at a redshift of $z\approx4.22$,
obtained with the {\it Mid-IR Instrument} (MIRI)
on board JWST (Spilker et al.\ 2023).
The 3.3 and 3.4$\mum$ emission features
have also been detected by JWST,
through its {\it Near Infrared Camera} (NIRCam),
in dozens of moderately distant galaxies
at redshifts $z$\,$\sim$\,0.2--0.5
in the {\it Great Observatories Origins
Deep Survey--South}
(GOODS-S; see Lyu et al.\ 2023).
It is expected that JWST will accumulate
a rich set of such spectra for a wide range of
astrophysical regions,
particularly in the distant universe.

We have initiated a program to explore
the aliphatic and aromatic contents
of PAHs in the universe, both in the
Milky Way and external galaxies,
both near and far.
In this work, we focus on the 3.3 and
3.4$\mum$ emission features detected
in the pre-JWST era.
This paper is organized as follows.
In \S\ref{sec:model} we present
a theoretical framework for
relating the aliphatic fractions of PAHs
to the emission intensity ratios of
the 3.4$\mum$ feature to the 3.3$\mum$ feature.
This theoretical framework will not only be used
in later sections but in the very near future also
serve the JWST community as an effective tool
for quantitatively determining
the aliphatic fractions of PAHs.
The 3.3 and 3.4$\mum$ emission features
of various astrophysical regions
detected in the pre-JWST era will be summarized
and analyzed in \S\ref{sec:obs}.
We will quantitatively determine the aliphatic fractions
of PAHs and discuss the results in \S\ref{sec:discussion}.
Finally, we summarize our major results
in \S\ref{sec:summary}.

\section{IR Emission Spectra of PAHs with
            Aliphatic Sidegroups:
            Theoretical Framework}
               \label{sec:model}
To facilitate the analysis of the 3.3 and 3.4$\mum$
emission detected in the pre-JWST era, we first set up
a theoretical framework to model the IR emission of
PAHs containing aliphatic sidegroups and relate
the emission intensities of the 3.3 and 3.4$\mum$
features to the PAH aliphatic fraction.
In the JWST era, this theoretical framework will
also be used to analyze JWST observations
to quantitatively determine the aliphatic fractions of PAHs.

Due to their small heat contents,
PAHs are transiently heated in the ISM
by single stellar photons (see Li 2004).
They will not attain an equilibrium temperature,
instead, they will experience temperature spikes
and undergo temperature fluctuations.
For PAHs containing aliphatic contents
(which we call ``aliphatic'' PAHs),
we consider PAHs attached with aliphatic
sidegroups like methylene and methyl.
Following Draine \& Li (2001), we will calculate
the temperature probability distribution functions
and emission spectra of aliphatic PAHs of
$\NCaro$ aromatic C atoms,
$\NHaro$ aromatic H atoms,
$\NCali$ aliphatic C atoms,
and $\NHali$ aliphatic H atoms.
For such molecules, we approximate
their absorption cross sections by
adding three Drude functions to that
of PAHs of $\NCaro$ C atoms
and $\NHaro$ H atoms
These Drude functions represent
the 3.4$\mum$ aliphatic C--H stretch,
and the 6.85 and 7.25$\mum$ aliphatic
C--H deformations. The absorption cross section
of an aliphatic PAH molecule of $\NCaro$ aromatic
C atoms, $\NHaro$ aromatic H atoms,
$\NCali$ aliphatic C atoms,
and $\NHali$ aliphatic H atoms becomes
\begin{eqnarray}
\label{eq:Cabs1}
C_{\rm abs}(\NC,\lambda) & = & \CabsPAH(\NCaro,\NHaro,\lambda)\\
\label{eq:Cabs2}
& + & \NHali \frac{2}{\pi}
    \frac{\gamma_{3.4} \lambda_{3.4} \sigma_{\rm int,3.3}
     \left(A_{3.4}/A_{3.3}\right)}
     {(\lambda/\lambda_{3.4}-\lambda_{3.4}/\lambda)^2
      +\gamma_{3.4}^2}\\
\label{eq:Cabs3}
&+& \NHali \frac{2}{\pi}
     \frac{\gamma_{6.85} \lambda_{6.85}
     \sigma_{\rm int,6.2} \left(\Adfa/\Acc\right)}
     {(\lambda/\lambda_{6.85}-\lambda_{6.85}/\lambda)^2
    +\gamma_{6.85}^2}\\
\label{eq:Cabs4}
&+& \NHali \frac{2}{\pi}
    \frac{\gamma_{7.25} \lambda_{7.25}
    \sigma_{\rm int,6.2} \left(\Adfb/\Acc\right)}
     {(\lambda/\lambda_{7.25}-\lambda_{7.25}/\lambda)^2
    +\gamma_{7.25}^2} ~~,
\end{eqnarray}
where $\NC=\NCaro+\NCali$ is the number of C atoms
contained in an aliphatic PAH molecule;
$\lambda_{3.4}=3.4\mum$,
$\lambda_{6.85}=6.85\mum$,
and $\lambda_{7.25}=7.25\mum$
are respectively the central wavelengths of
the 3.4, 6.85 and 7.25$\mum$ aliphatic C--H features;
$\gamma_{3.4}\lambda_{3.4}$,
$\gamma_{6.85}\lambda_{6.85}$,
and $\gamma_{7.25}\lambda_{7.25}$
are respectively the FWHMs
of the 3.4, 6.85 and 7.25$\mum$ features
($\gamma_{3.4}$, $\gamma_{6.85}$,
and $\gamma_{7.25}$ are dimentionless parameters;
see Draine \& Li 2007);
$A_{3.3}$ and $A_{3.4}$ are the intensities of the
aromatic and aliphatic C--H stretches, respectively;
$A_{6.2}$ and $A_{7.7}$ are the intensities of
the C--C stretches;
$A_{6.85}$ and $A_{7.25}$ are the intensities of
the aliphatic C--H deformation bands;
and $\sigma_{{\rm int},3.3}$ and $\sigma_{{\rm int},6.2}$
are respectively the integrated strengths per (aromatic)
C atom of the 3.3$\mum$ aromatic C--H stretch
and 6.2$\mum$ aromatic C--C stretch
(see Draine \& Li 2007).
We take $A_{3.4}/A_{3.3}=1.76$ for neutrals
and $A_{3.4}/A_{3.3}=3.80$ for cations
as computed by Yang et al.\ (2013).
We take the lower limits of $\Adfa/\Acc\approx5.0$
and $\Adfb/\Acc\approx0.5$ for neutrals,
$\Adfa/\Acc\approx0.5$ and $\Adfb/\Acc\approx0.25$
for cations as derived in Yang et al. (2016a).
%
We note that, with $\NCali\approx3\NHali$
(suitable for methyl sidegroups),
the absorption cross sections
given in eqs.\ref{eq:Cabs1}--\ref{eq:Cabs4}
are the same as that of Yang et al.\ (2016a).
%

Let $dP$ be the probability that the temperature
of the aliphatic PAH molecule will be in $[T,T+dT]$.
The emissivity (in unit of $\erg\s^{-1}\cm^{-1}$)
of this molecule becomes
\begin{equation}
j_\lambda(\NC) = \int C_{\rm abs}(\NC,\lambda)\,
            4\pi B_\lambda(T)\,\frac{dP}{dT}\,dT  ~.
\end{equation}
The 3--4$\mum$ interstellar UIE emitters are
in the size range of $\NC$\,$\simali$20--30 C atoms,
as shown in Figures~6, 7 of Draine \& Li (2007).
For illustrative purpose, we consider $\NCaro=24$
(like coronene). For a coronene-like molecule,
up to 12 methylene or methyl sidegroups
can be attached, we thus consider
$\NCali=0, 1, 2, ...12$ aliphatic C atoms
and $\NHali=0, 1, 2, ...36$ aliphatic H atoms.
For all molecules, $\NCaro=24$ is fixed.
Yang et al.\ (2016a) have shown that the model
IR emission spectra (scaled by starlight intensity)
are essentially independent of the absolute values
of the starlight intensities.
Therefore, we only consider $U=1$,
with $U$ defined as
\begin{equation}
U \equiv \frac{\int_{1\mu {\rm m}}^{912{\rm \Angstrom}}
               4\pi J_\star(\lambda)\,d\lambda}
              {\int_{1\mu {\rm m}}^{912{\rm \Angstrom}}
               4\pi J_{\rm ISRF}(\lambda)\,d\lambda} ~~,
\end{equation}
where $J_\star(\lambda)$ is the intensity of starlight,
and $J_{\rm ISRF}(\lambda)$ is the starlight intensity of
the solar neighbourhood interstellar radiation field (ISRF)
of Mathis, Mezger \& Panagia (1983; MMP83).
In addition to the MMP83 ISRF,
we consider five types of radiation fields,
approximated by the stellar model atmospheric
spectra of Kurucz (1979) of effective temperatures
of $\Teff=3,500, 6,000, 10,000, 22,000, 30,000\K$,
like that of M2V stars, the Sun, A2V stars, B1.5V stars
and B0V stars, respectively. The reflection nebula
NGC~2023 is illuminated by HD~37903,
an B1.5V star with $\Teff=22,000\K$,
while IRAS~03035+5819 is illuminated by
an B0V star of $\Teff=30,000\K$.

\begin{figure*}
\centering{
\includegraphics[scale=0.4,clip]{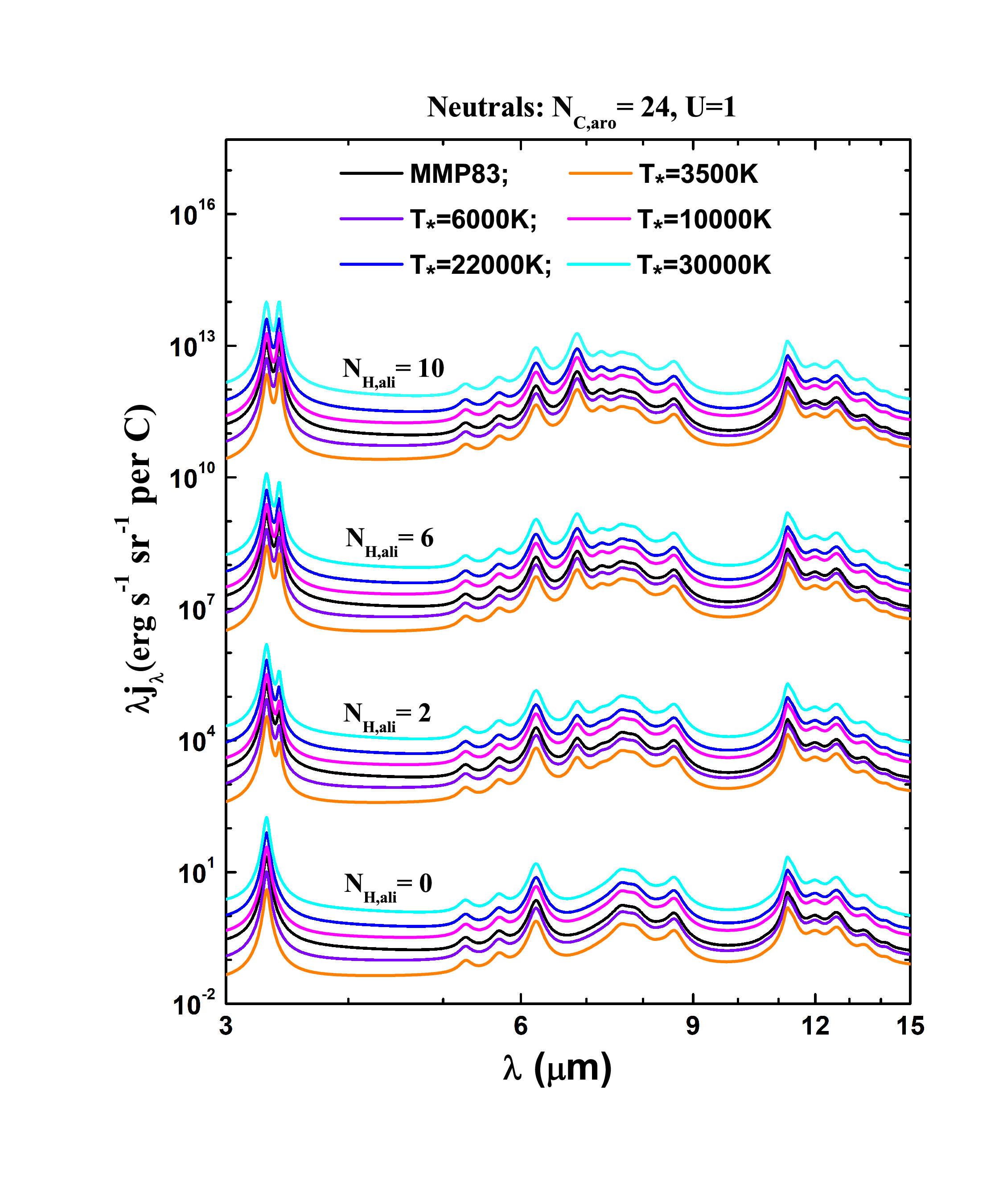}
}
\caption{\footnotesize
         \label{fig:Spec_nPAH_BR}
         Model IR emission spectra of neutral aliphatic PAHs
         of $\NHali=0, 2, 6, 10$ aliphatic H atoms
         and $\NCaro=24$ aromatic C atoms
         illuminated by an M2V star of
         $\Teff=3,500\K$ (orange lines),
         a solar-type star of
         $\Teff=6,000\K$ (purple lines),
         an A2V star of
         $\Teff=10,000\K$ (mangeta lines),
         an B1.5V star of $\Teff=22,000\K$ (blue lines),
         an B0V star of $\Teff=30,000\K$ (cyan lines),
         and the MMP83 ISRF (black lines).
         The starlight intensities are all set
         to be $U=1$.
         The 3.4 and 6.85$\mum$ aliphatic C--H features
         are clearly seen in the spectra of aliphatic PAHs
         with $\NHali=2, 6, 10$, while the 7.25$\mum$ aliphatic
         C--H feature is less prominent.
         For clarity, their spectra
         are vertically shifted.
         }
\end{figure*}

\begin{figure*}
\centering{
\includegraphics[scale=0.4,clip]{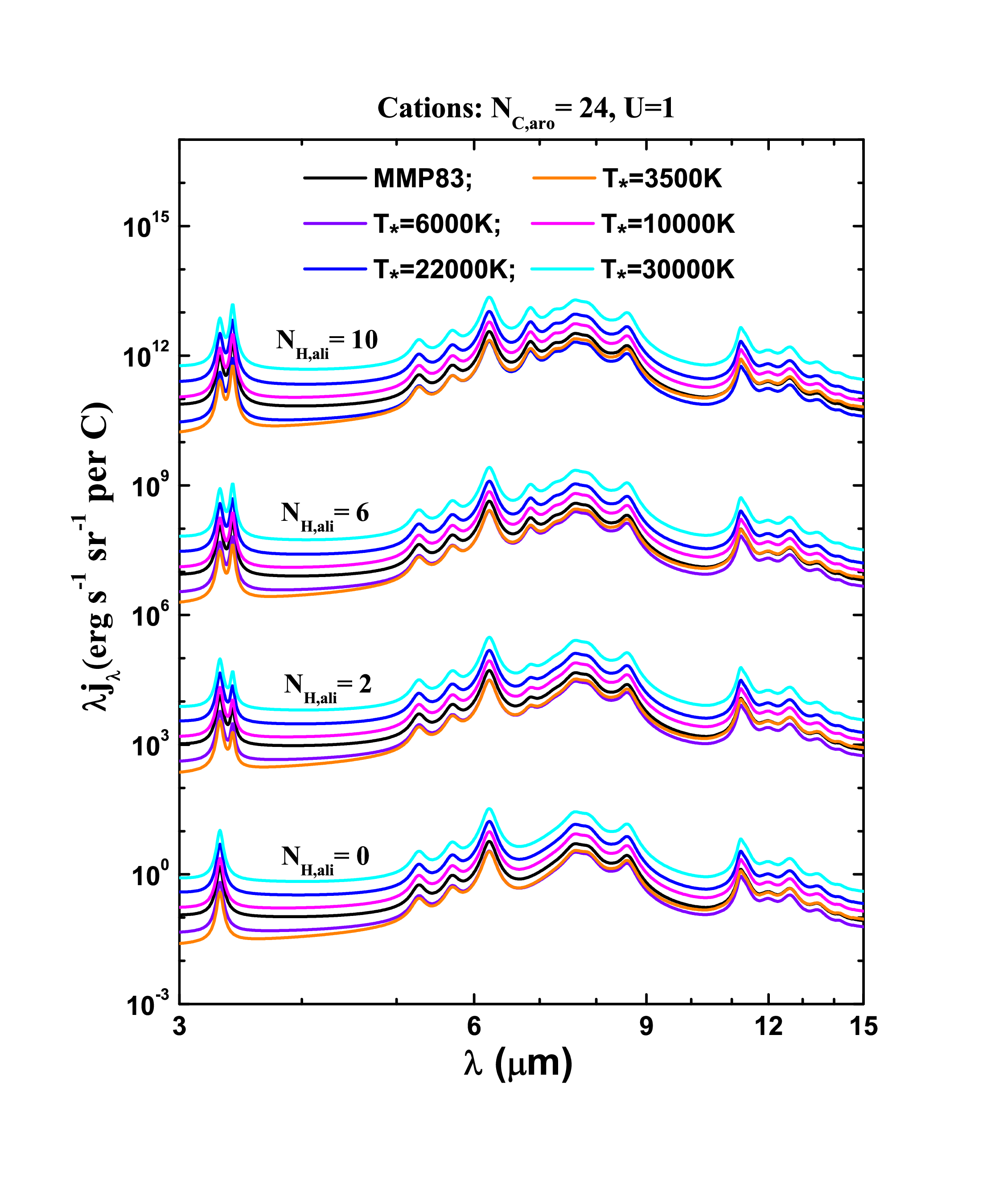}
}
\caption{\footnotesize
         \label{fig:Spec_iPAH_BR}
         Same as Figure~\ref{fig:Spec_nPAH_BR}
         but for aliphatic PAH cations.
         }
\end{figure*}

We adopt the ``thermal-discrete'' method
of Draine \& Li (2001) to compute the temperature
probability distribution functions
and the IR emission spectra of both neutral
and ionized aliphatic PAHs
excited by starlight of different spectra.
In Figures~\ref{fig:Spec_nPAH_BR} and \ref{fig:Spec_iPAH_BR}
we show the model emission spectra in 3--15$\mum$
respectively for neutral and ionized aliphatic PAHs
of $\NHali=0, 2, 6, 10$
illuminated by stars of different $\Teff$.
It is apparent that the 3.4 and 6.85$\mum$
aliphatic C--H features are clearly visible
and become stronger as $\NHali$ increases.
The 7.25$\mum$ aliphatic C--H feature,
however, remains hardly noticeable even
for $\NHali=10$. This is because the intrinsic
strength of the 7.25$\mum$ feature ($A_{7.25}$)
is much weaker compared to that of the 3.4 and
6.85$\mum$ features  ($A_{3.4}$, $A_{6.85}$;
see Yang et al.\ 2016a).

\begin{figure*}
\centering{
\includegraphics[scale=0.4,clip]{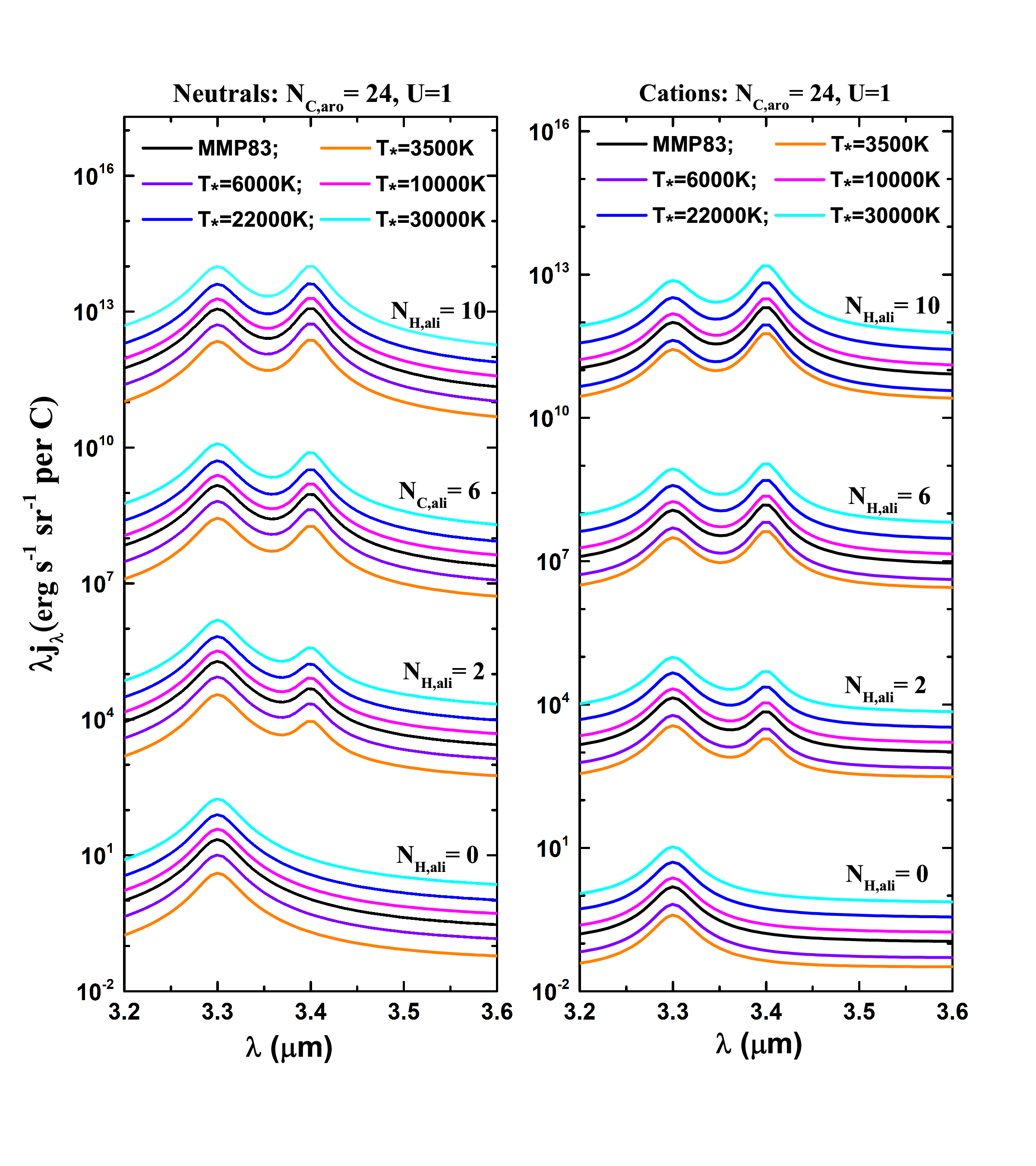}
}
\caption{\footnotesize
         \label{fig:Spec_inPAH_SR}
         Same as Figures~\ref{fig:Spec_nPAH_BR},\ref{fig:Spec_iPAH_BR}
         but highlighting the aromatic and aliphatic
         C--H bands in the 3.2--3.6$\mum$
         wavelength range.
          }
\end{figure*}

In the following, we will focus on
the 3.3 and 3.4$\mum$ features.
Figure~\ref{fig:Spec_inPAH_SR} highlights
the spectra in the wavelength range of
3.2--3.6$\mum$ for both neutral
and ionized aliphatic PAHs with
$\NHali=0, 2, 6, 10$.
The 3.4$\mum$ aliphatic C--H band
becomes pronounced even at $\NHali=2$.
At $\NHali=10$, the 3.4$\mum$ feature
becomes comparable to or even stronger
than the 3.3$\mum$ aromatic C--H feature.
For the same $\NHali$, PAH cations emit
less at 3.3 and 3.4$\mum$ than their
neutral counterparts.

For a given $\NHali$, we derive $\Iratiomod$,
the model emission intensity ratio of
the 3.4$\mum$ band to the 3.3$\mum$ band, from
\begin{equation}\label{eq:Iratiomod}
\left(\frac{I_{3.4}}{I_{3.3}}\right)_{\rm mod}
= \frac{\int_{3.4}\Delta j_\lambda(\NC)\,d\lambda}
{\int_{3.3}\Delta j_\lambda(\NC)\,d\lambda} ~~,
\end{equation}
where $I_{3.4}$ and $I_{3.3}$ are respectively
the calculated intensities of the 3.4$\mum$
and 3.3$\mum$ emission features; and
$\int_{3.3}\Delta j_\lambda(\NC)\,d\lambda$
and $\int_{3.4}\Delta j_\lambda(\NC)\,d\lambda$
are respectively the feature-integrated excess emission
of the 3.3 and 3.4$\mum$ features of aliphatic PAHs.

\begin{figure*}
\centering{
\includegraphics[scale=0.5,clip]{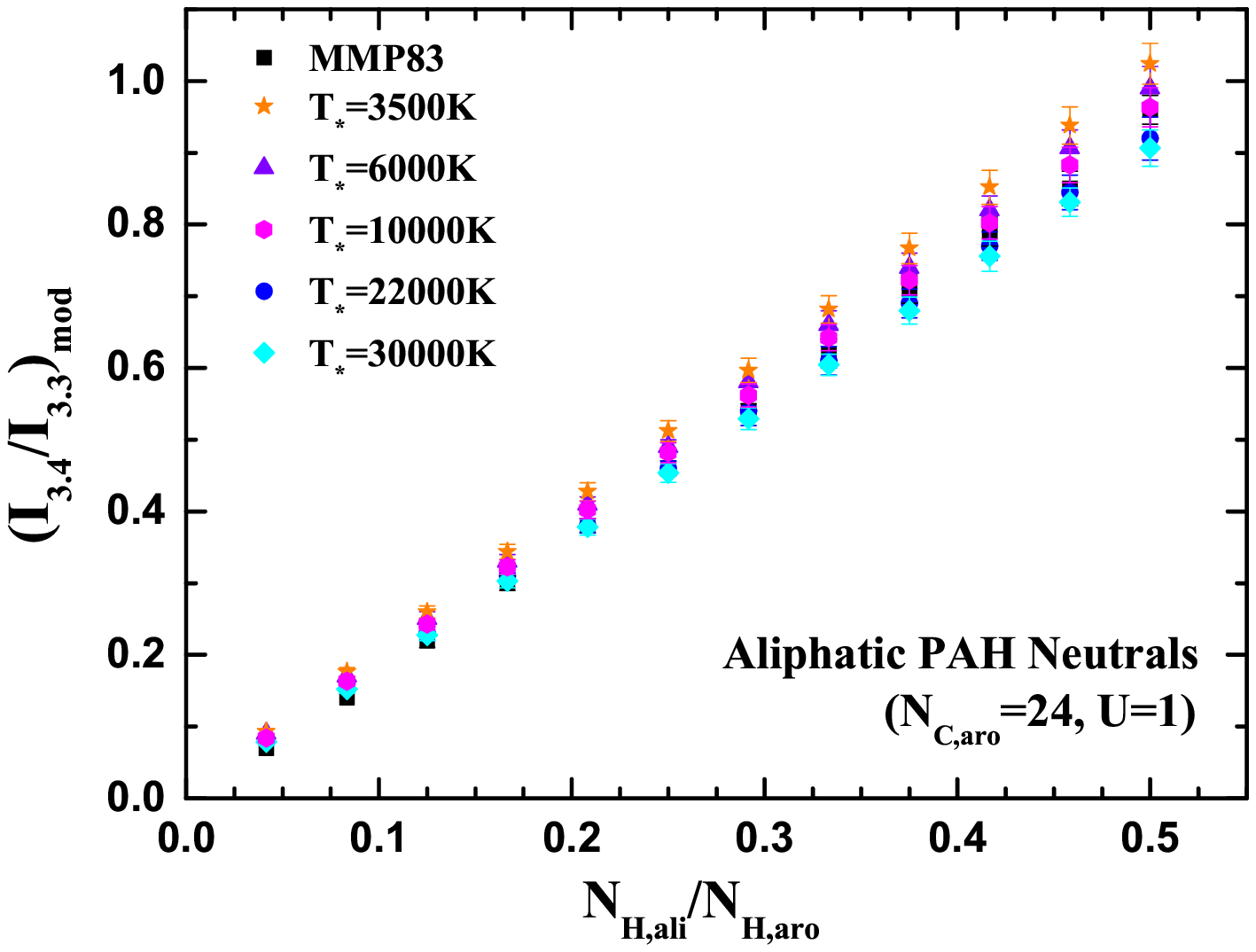}
}
\caption{\footnotesize
             \label{fig:nPAH_Iratio_NC24_T}
             Model-calculated intensity ratios $\Iratiomod$
              as a function of $\NHali/\NHaro$
              for neutral aliphatic PAHs of $\NCaro=24$.
              These molecules are illuminated by
              an M2V star of $\Teff=3,500\K$ (orange lines),
         a solar-type star of $\Teff=6,000\K$ (purple lines),
         an A2V star of $\Teff=10,000\K$ (magenta lines),
         an B1.5V star of $\Teff=22,000\K$ (blue lines),
         an B0V star of $\Teff=30,000\K$ (cyan lines),
         and the MMP83 ISRF (black lines).
             The starlight intensities are all set
             to be $U=1$.
          }
\end{figure*}

\begin{figure*}
\centering{
\includegraphics[scale=0.5,clip]{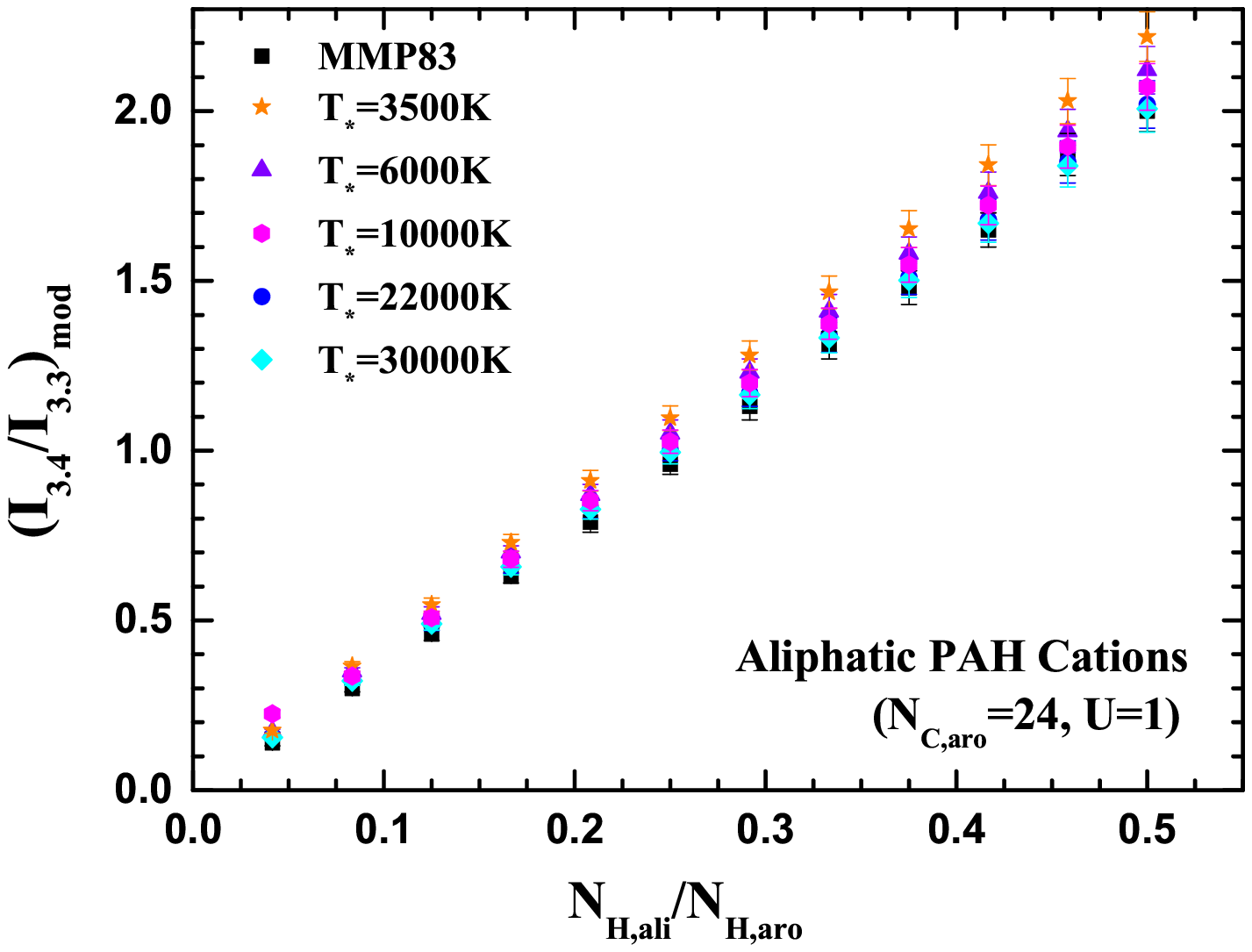}
}
\caption{\footnotesize
         \label{fig:iPAH_Iratio_NC24_T}
         Same as Figure~\ref{fig:nPAH_Iratio_NC24_T}
         but for aliphatic PAH cations.
         }
\end{figure*}

In Figures~\ref{fig:nPAH_Iratio_NC24_T}
and \ref{fig:iPAH_Iratio_NC24_T}
we show the model intensity ratios $\Iratiomod$
as a function of $\NHali/\NHaro$ for
neutral and ionized PAHs, respectively.
Basically, the model band ratios $\Iratiomod$
are linearly correlated with $\NHali/\NHaro$
for both neutrals and cations.
The correlation slope, defined as
$d\Iratiomod/d\left(\NHali/\NHaro\right)$,
is a weak function of $\Teff$
and listed in Table~\ref{tab:Slope-all}.
On average,
$\langle d\Iratiomod/d\left(\NHali/\NHaro\right)\rangle
\approx1.92\pm0.09$ for neutrals
and $\approx4.13\pm0.16$ for cations.
Therefore, to first order, we obtain
$\Iratiomod\approx1.92\times \left(\NHali/\NHaro\right)$ for neutrals
and $\Iratiomod\approx4.13\times \left(\NHali/\NHaro\right)$ for cations.
With the temperature dependence of
the correlation slope taken into account,
the model band ratio $\Iratiomod$
can be expressed as
\begin{equation}\label{eq:Iratiomod}
\left(\frac{I_{3.4}}{I_{3.3}}\right)_{\rm mod}
= \left(\frac{A_{3.4}}{A_{3.3}}\right)
\times\left(\frac{\NHali}{\NHaro}\right)\times k(\Teff) ~~,
\end{equation}
where $k(\Teff)$, the correlation slope, is
\begin{equation}\label{eq:slope}
k(\Teff)\approx
\begin{cases}
1.20-0.122\times\left(\Teff/10,000\K\right)
+0.022\times\left(\Teff/10,000\K\right)^2
& {\rm for~neutrals} ~~,\\
1.18-0.113\times\left(\Teff/10,000\K\right)
+0.023\times\left(\Teff/10,000\K\right)^2
& {\rm for~cations} ~~.\\
\end{cases}
\end{equation}
The correlation slope $k(\Teff)$
somewhat decreases as $\Teff$ increases.
This is because, in regions illuminated by
hot stars (of higher $\Teff$), the stellar photons
are more energetic. Upon absorption of
such an energetic photon emitted from hotter stars,
PAHs are excited to higher temperatures and
emit more effectively at shorter wavelengths
(e.g., 3.3$\mum$) than at longer wavelengths
(e.g., 3.4$\mum$). Therefore, for a given
$\NHali/\NHaro$, a smaller $\Iratiomod$
is expected for regions illuminated by stars
of higher $\Teff$.

To relate $\NCali/\NCaro$ through $\NHali/\NHaro$,
we assume that one aliphatic C atom
corresponds to 2.5 aliphatic C--H bonds
(intermediate between methylene --CH$_2$
and methyl --CH$_3$)
and one aromatic C atom corresponds to
0.75 aromatic C--H bond
(intermediate between benzene C$_6$H$_6$
and coronene C$_{24}$H$_{12}$).
Therefore, the ratio of the number of C atoms
in aliphatic units to that in aromatic rings is
$\NCali/\NCaro\approx
\left(0.75/2.5\right)\,\times\,\NHali/\NHaro$.
As the 3.3 and 3.4$\mum$ C--H stretches
are predominantly emitted by neutral PAHs,
we therefore recommend the following relation
to estimate $\NCali/\NCaro$
from the observed  band ratios $\Iratioobs$:
\begin{equation}\label{eq:alifrac1}
\frac{\NCali}{\NCaro}
\approx \frac{1}{5.87}
\left(\frac{I_{3.4}}{I_{3.3}}\right)_{\rm obs}
\times\left\{1.20-0.122\times\left(\Teff/10,000\K\right)
+0.022\times\left(\Teff/10,000\K\right)^2 \right\}^{-1} ~~.
\end{equation}
In case there is no information on
$\Teff$ (e.g., the MMP83 ISRF),
we recommend
\begin{equation}\label{eq:alifrac2}
\frac{\NCali}{\NCaro}
\approx \frac{1}{6.40}
\left(\frac{I_{3.4}}{I_{3.3}}\right)_{\rm obs}~~.
\end{equation}
%
%
The aliphatic fraction of PAHs
is determined from
\begin{equation}\label{eq:alifrac3}
\alifrac = \left(1+\NCaro/\NCali\right)^{-1} ~~.
\end{equation}
There is no need to compute
the temperature probability
distribution functions and
the IR emission spectra of
aliphatic PAHs as long as one
is only interested in the aliphatic
fraction of the UIE carrier.

\section{Aliphatic and Aromatic Observations
            in the Pre-JWST Era}\label{sec:obs}
A wealth of observational spectra for the aliphatic and
aromatic C--H stretches are available in archive or literature.
This allows an in-depth study of the aliphatics and aromatics
in the universe. We compile the aliphatic and aromatic C--H
emission data, as complete as possible, from observations
made with space-borne satellites
such as the {\it Infrared Space Observatory} (ISO) and AKARI,
airborne telescopes such as the {\it Kuiper Airborne Observatory}
(KAO), and ground-based telescopes
such as the {\it Infrared Telescope Facilities} (IRTF)
and the {\it United Kingdom Infrared Telescope} (UKIRT).

To this end, we find 28 sources which show
both the 3.3 and 3.4$\mum$ features.
These sources include Galactic PDRs,
protoplanetary nebulae (PPNe),
planetary nebulae (PNe),
reflection nebulae (RNe),
young stellar objects (YSOs),
and HII regions, as well as external galaxies.

For each source, we fit the observed spectrum
in terms of two or more Drude profiles combined
with an underlying linear continuum:
\begin{equation}\label{eq:drude}
F_\lambda = a_0 + a_1 \lambda + \sum_j
\frac{P_j\,\times\,\left(2\gamma_j/\pi\right)}
{\left(\lambda-\lambda_{{\rm o},j}^2/\lambda\right)^2+\gamma_j^2},
\end{equation}
%
where $a_0$ and $a_1$ are the coefficients
of the linear continuum; $\lambda_{{\rm o},j}$
and $\gamma_j$ are the central wavelength
and width of the $j$-th Drude profile;
$P_j$, the power emitted from $j$-th Drude profile
(in unit of $\erg\s^{-1}\cm^{-2}$),
is obtained by integrating the emission feature
over wavelength:
\begin{equation}\label{eq:P_j}
P_j = \int_{\lambda_j}\Delta F_\lambda\,d\lambda ~~.
\end{equation}
%
%
For the Drude profiles, the 3.3 and 3.4$\mum$ features
are always included for consideration.
In some objects (e.g., IRAS\,21282+5050),
one or more additional weak features
at 3.43, 3.47, 3.51, and 3.56$\mum$
are also present and each of these features
is also approximated as a Drude profile.
We sum up the power emitted from all
these sub-features and attribute them
to the aliphatic C--H stretches.
Therefore, for the ratio of the power
emitted from the aliphatic C--H stretches
to that from the aromatic C--H stretches,
we take
$\Iratioobs = \left(P_{3.4}+P_{3.43}+P_{3.47}
+P_{3.51}+P_{3.56}\right)/P_{3.3}$
provided that these subfeatures are detected.
If only the 3.3 and 3.4$\mum$ features
show up, we take $\Iratioobs = P_{3.4}/P_{3.3}$.
We note that in the literature the band ratios
$\Iratioobs$ have been reported for some sources.
We prefer to  derive by ourselves because
in the literature there is a certain arbitrarity
in defining the underlying continuum
for the features and the strengths
of the features were calculated in different ways.
When taking data from different publications,
these differences may actually play a role.
We therefore decide to derive $\Iratioobs$
in a coherent way for all sources.

\begin{figure*}
\centering{
\includegraphics[scale=0.4,clip]{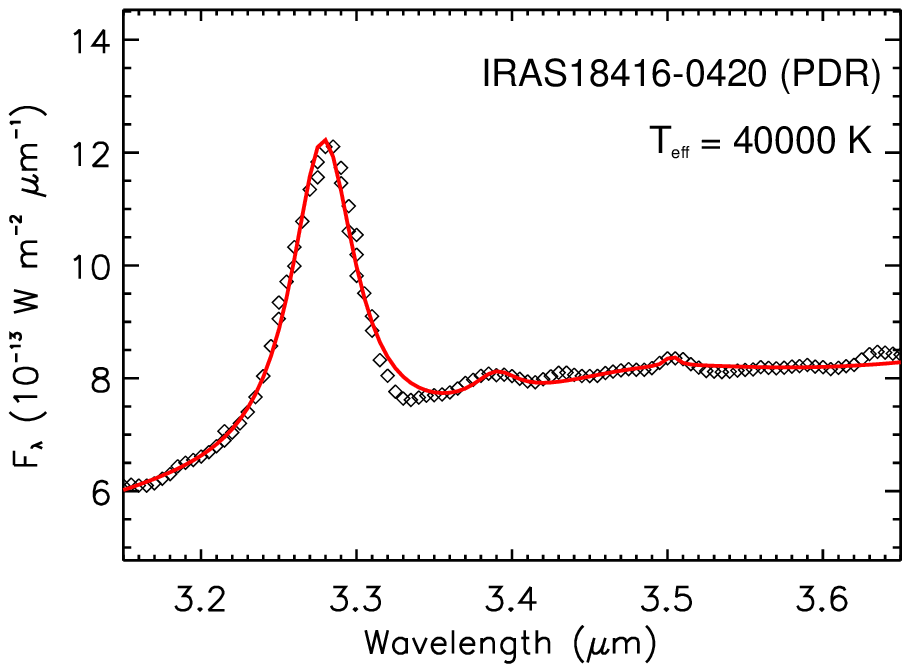}
\includegraphics[scale=0.4,clip]{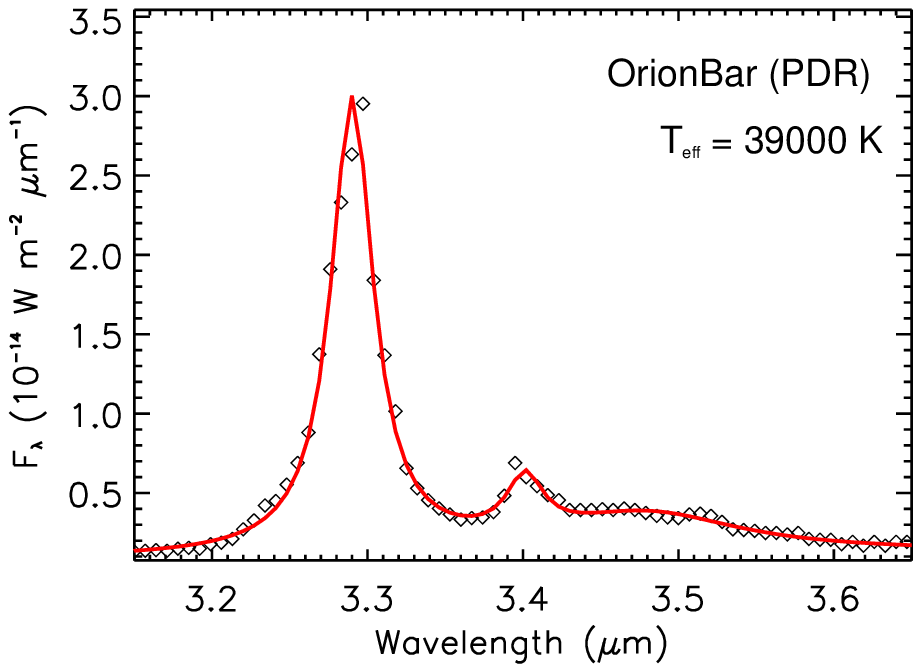}
\includegraphics[scale=0.4,clip]{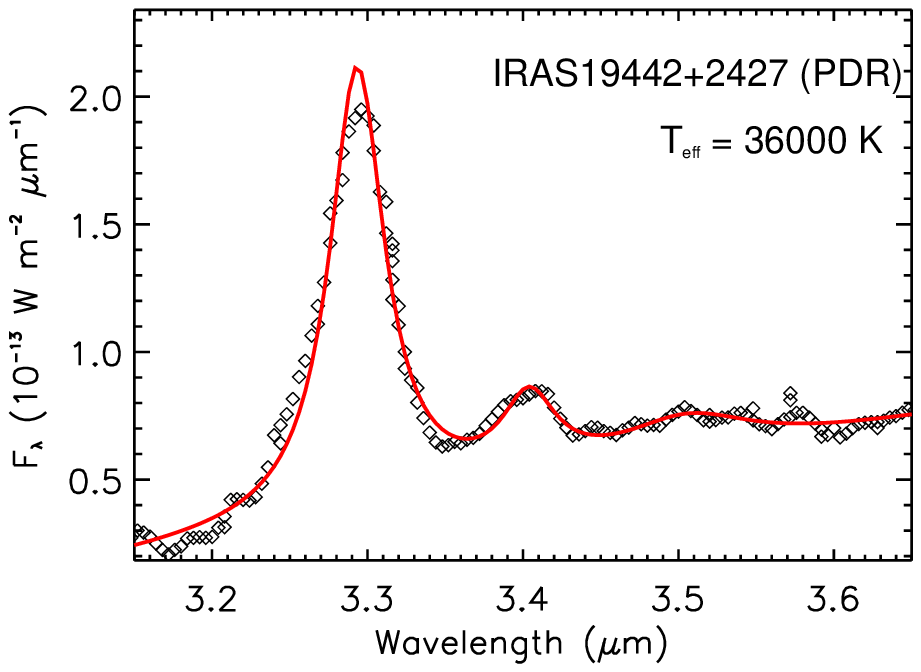}
\includegraphics[scale=0.4,clip]{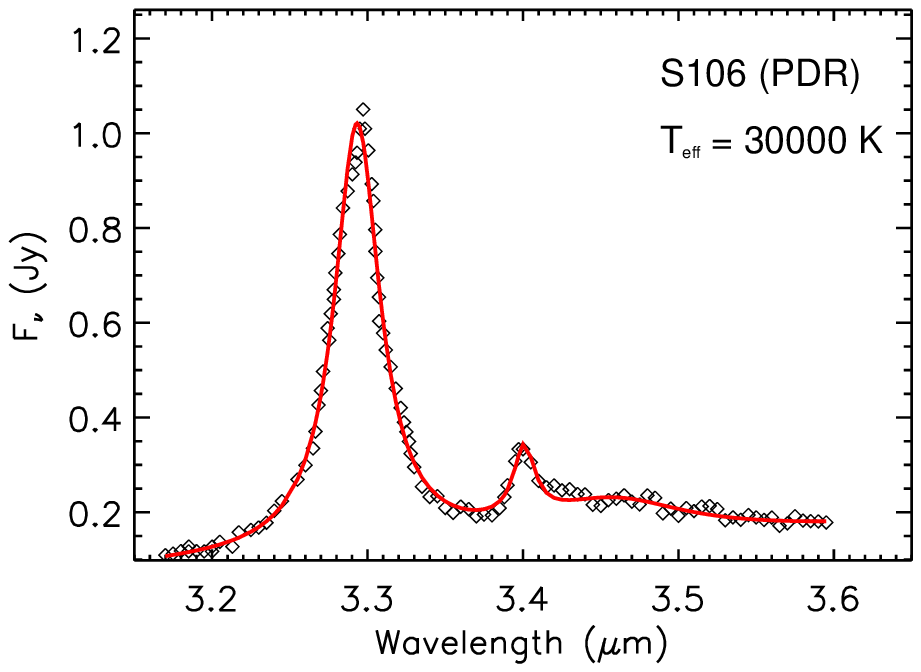}
}
\caption{\footnotesize
         \label{fig:Spec_PDR}
         Aliphatic and aromatic C--H
         stretching features of four PDRs
         (IRAS~18416-0420, Jourdain de Muizon et al.\ 1990;
         Orion Bar, Sloan et al.\ 1997;
         IRAS~19442+2427, Jourdain de Muizon et al.\ 1990;
         S106, Geballe et al.\ 1985).
         The observed spectra are shown as black diamonds.
         The fitted spectra (solid red lines) are a combination
         of two or more Drude profiles and a linear, underlying
         continuum.
          }
\end{figure*}

\begin{figure*}
\centering{
\includegraphics[scale=0.4,clip]{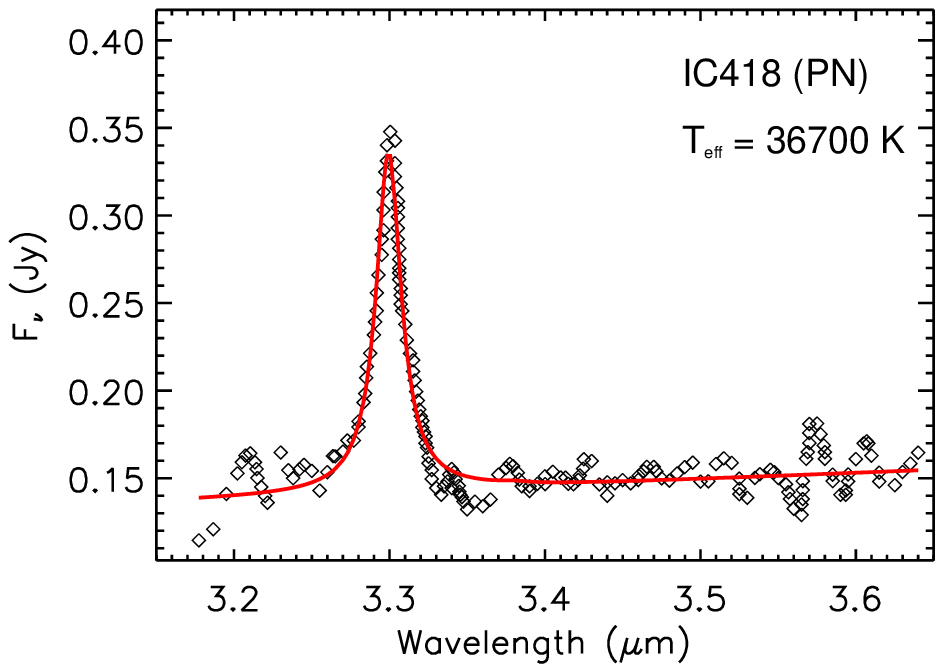}
\includegraphics[scale=0.4,clip]{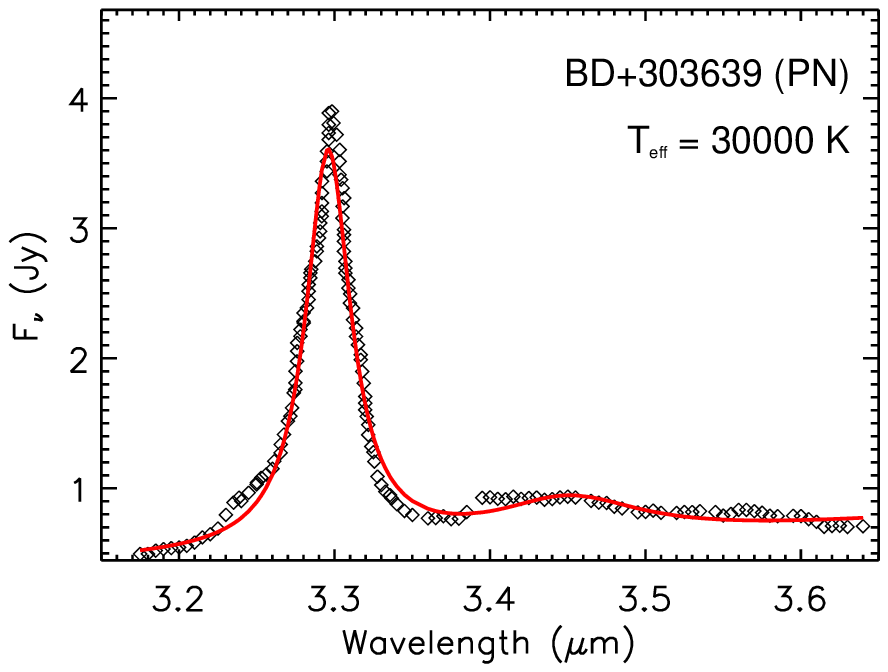}
\includegraphics[scale=0.4,clip]{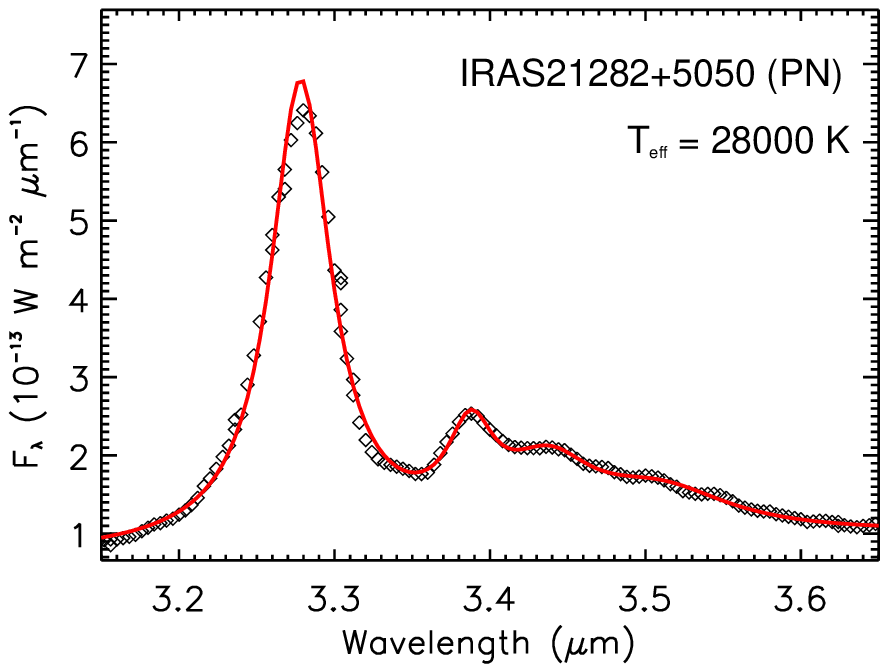}
\includegraphics[scale=0.4,clip]{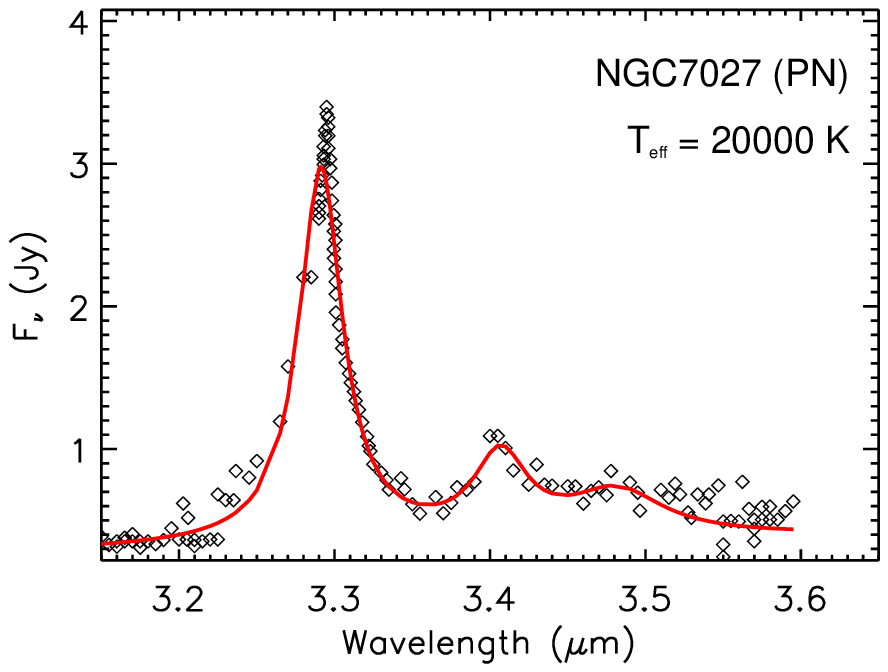}
}
\caption{\footnotesize
         \label{fig:Spec_PN}
         Same as Figure~\ref{fig:Spec_PDR}
         but for four planetary nebulae
         (IC~418, Geballe et al.\ 1985;
         BD+303639, Geballe et al.\ 1985;
         IRAS~21282+5050, Jourdain de Muizon et al.\ 1986;
         NGC~7027, Geballe et al.\ 1985).
          }
\end{figure*}

\begin{figure*}
\centering{
\includegraphics[scale=0.4,clip]{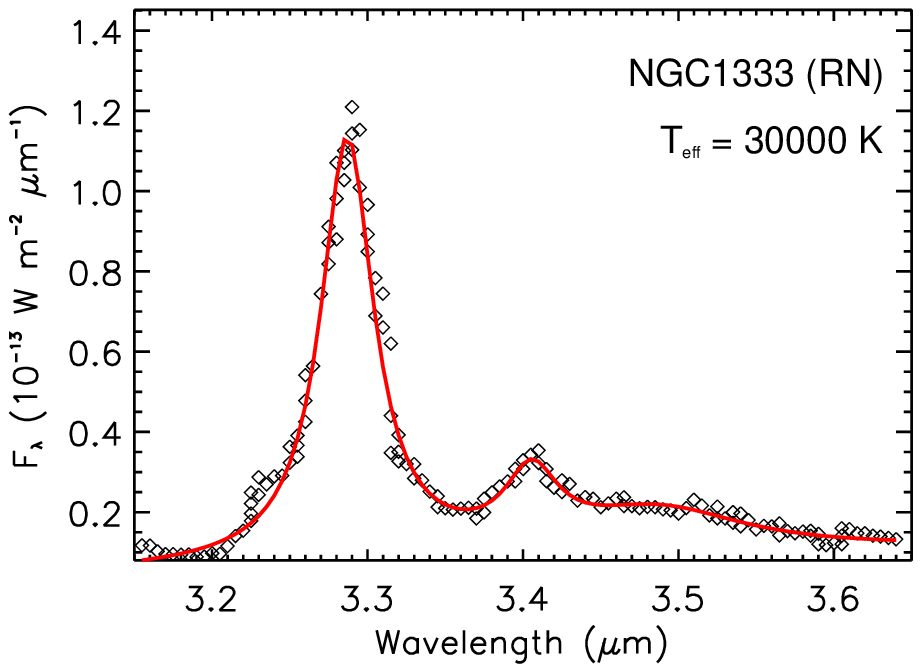}
\includegraphics[scale=0.4,clip]{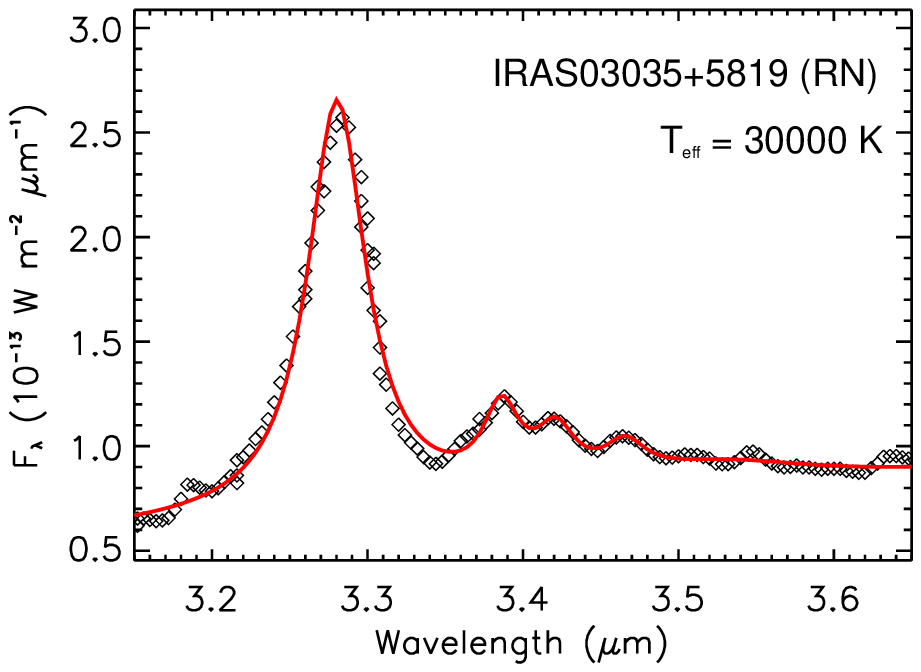}
}
\caption{\footnotesize
         \label{fig:Spec_RN}
         Same as Figure~\ref{fig:Spec_PDR}
         but for reflection nebulae
         NGC~1333 (Joblin et al.\ 1996)
         and IRAS~03035+5819
         (Jourdain de Muizon et al.\ 1986).
          }
\end{figure*}

\begin{figure*}
\centering{
\includegraphics[scale=0.4,clip]{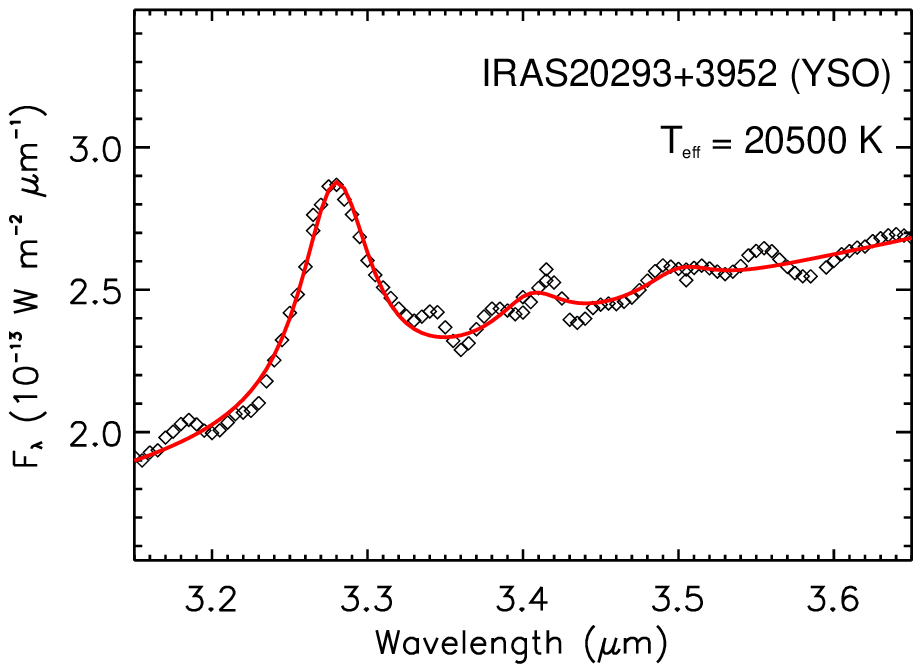}
\includegraphics[scale=0.4,clip]{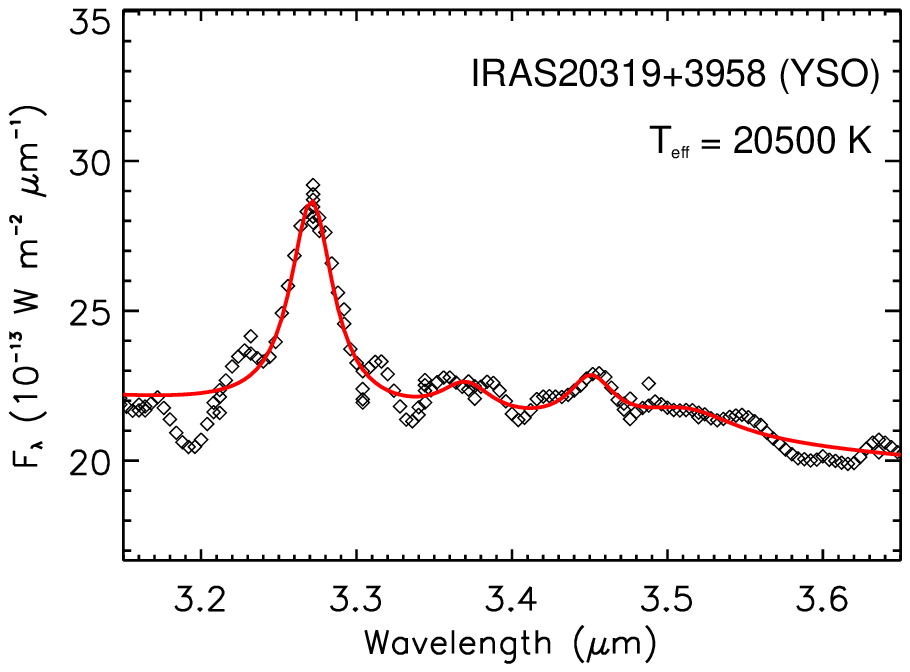}
\includegraphics[scale=0.4,clip]{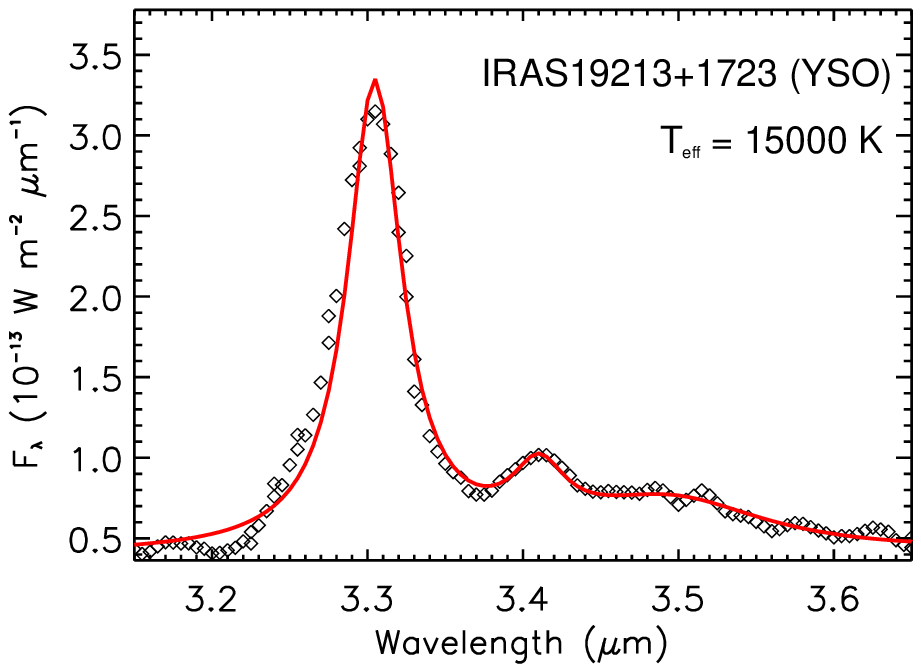}
\includegraphics[scale=0.4,clip]{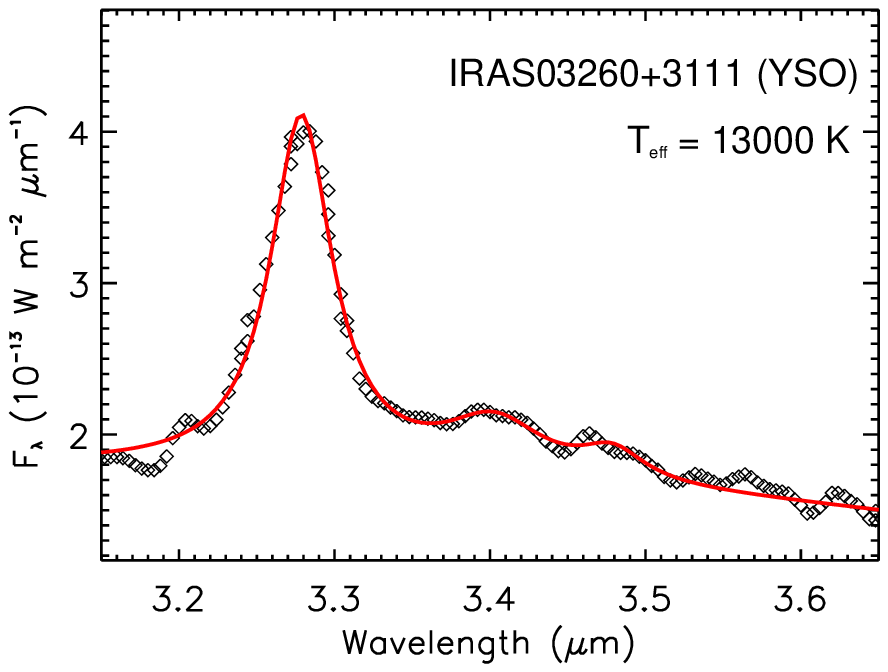}
}
\caption{\footnotesize
         \label{fig:Spec_YSO}
         Same as Figure~\ref{fig:Spec_PDR}
         but for four YSOs
         (IRAS~20293+3952,
         IRAS~20319+3958,
         IRAS~19213+1723,
         and IRAS~03260+3111;
         Jourdain de Muizon et al.\ 1990).
          }
\end{figure*}

For each source, we follow the above procedure
to fit the observed spectrum to derive $\Iratioobs$.
We then derive the aliphatic fraction $\alifrac$
from $\Iratioobs$
(see eqs.\,\ref{eq:alifrac1}--\ref{eq:alifrac3}).
The spectral fits are illustrated
in Figures~\ref{fig:Spec_PDR}--\ref{fig:Spec_PPN}
and the derived aliphatic fractions $\alifrac$
are tabulated in Table~\ref{tab:Source-aliphacity}.



\section{Results and Discussion}\label{sec:discussion}
Figure~\ref{fig:Spec_PDR} shows the aliphatic
and aromatic C--H stretches seen in emission
in PDRs excited by B0V or earlier-type stars
with $\Teff\simgt30,000\K$.
The aliphatic C--H stretches are relatively weak
and the aliphatic fractions of PAHs are all smaller
than 3\%. This is understandable since PDRs are
rich in energetic photons
so that the aliphatic sidegroups attached to PAHs
could easily be stripped off.

Figure~\ref{fig:Spec_PN} shows the near-IR spectra
of four PNe. NGC~7027, excited by an B2.5V star
of $\Teff\approx20,000\K$, exhibits the strongest
aliphatic C--H stretches among these four PNe.
As the illuminating star becomes hotter,
$\Iratioobs$ decreases,
so does the PAH aliphatic fraction.
With $\Teff\approx37,000\K$, IC~418 does not
show any noticeable aliphatic C--H emission.
In contrast, excited by an B0V star of
$\Teff\approx30,000\K$, BD+303639 shows
a broad, shallow feature around 3.4--3.5$\mum$.
By attributing this feature to aliphatic
C--H stretches, we estimate an aliphatic fraction
of $\simali$5.3\% for PAHs in BD+303639.

The aliphatic and aromatic C--H stretches of two
reflection nebulae are shown in Figure~\ref{fig:Spec_RN}.
IRAS~03035+5819 exhibits a series of
weak aliphatic C--H stretching features.
NGC~1333 and IRAS~03035+5819
are both excited by an B0V star of
$\Teff\approx30,000\K$ and their PAH
aliphatic fractions are comparable to
that of BD+303639 (PN), but appreciably
higher than that of S106 (PDR).
Both BD+303639 and S106 are illuminated
by stars with $\Teff\approx30,000\K$,
just like NGC~1333 and IRAS~03035+5819.

Figure~\ref{fig:Spec_YSO} shows
the near-IR spectra of four YSOs.
All four objects show several sub-features
at $\simali$3.4--3.6$\mum$
attributed to aliphatic C--H stretches.
The PAH aliphatic fraction does not show
any strong dependence on $\Teff$.

\begin{figure*}
\centering{
\includegraphics[scale=0.4,clip]{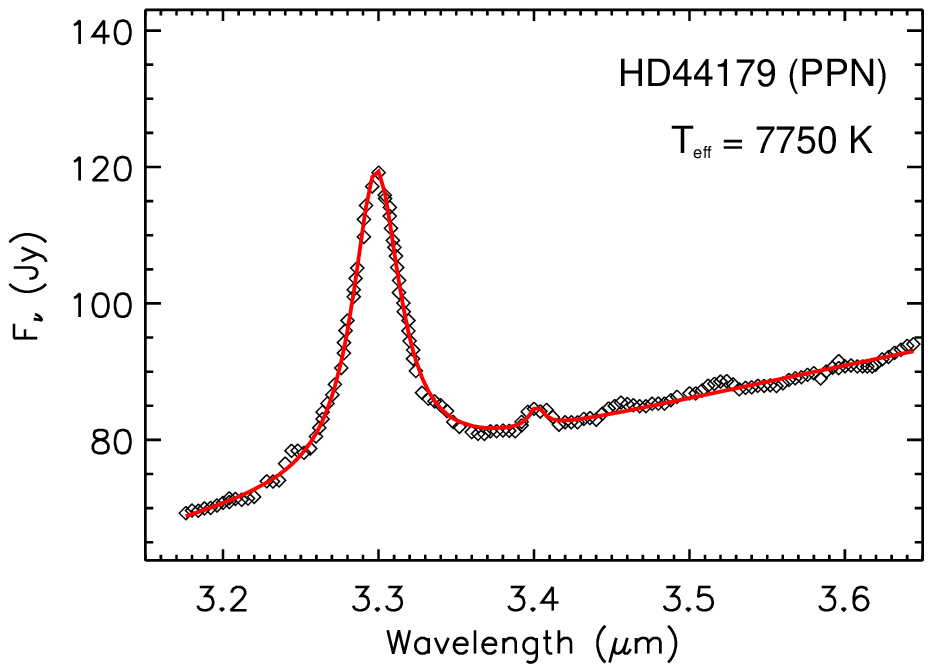}
\includegraphics[scale=0.4,clip]{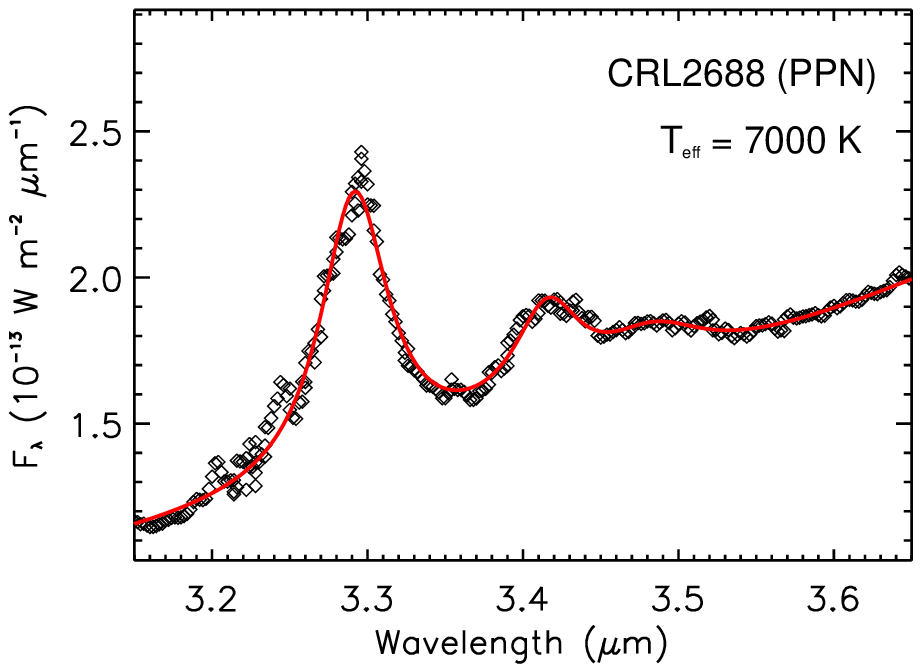}
\includegraphics[scale=0.4,clip]{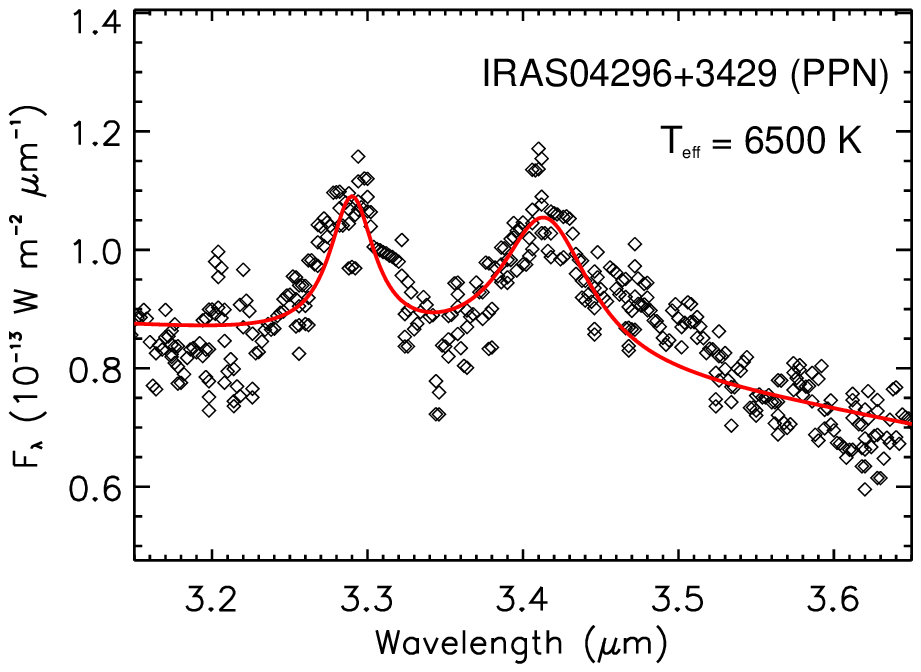}
\includegraphics[scale=0.4,clip]{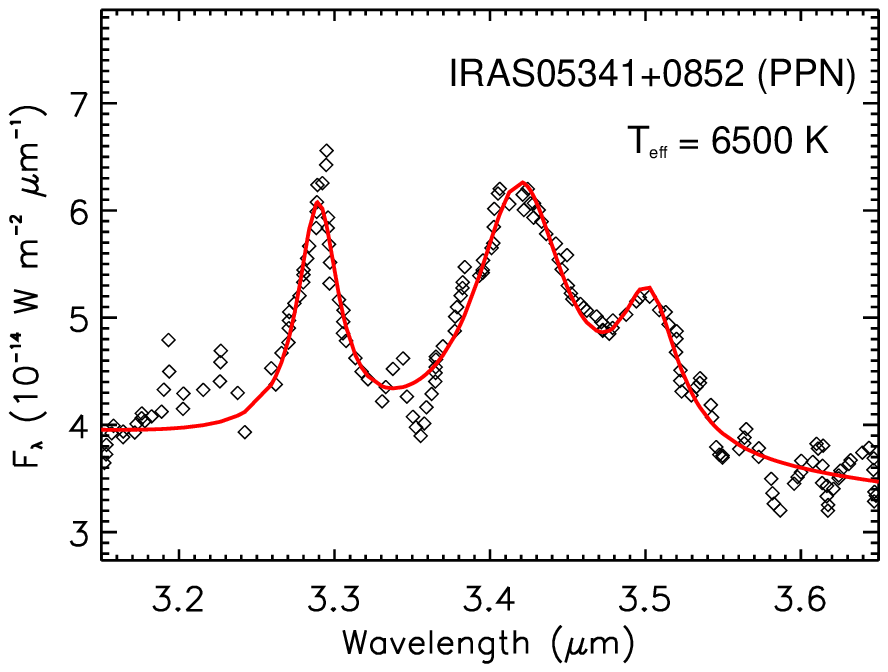}
}
\caption{\footnotesize
         \label{fig:Spec_PPN}
         Same as Figure~\ref{fig:Spec_PDR}
         but for four PPNe:
         the Red Rectangle illuminated
         by HD~44179 (Geballe et al.\ 1985);
         CRL~2688 (Geballe et al.\ 1992);
         IRAS~04296+3429 (Geballe et al.\ 1992);
         and IRAS~05341+0852 (Geballe et al.\ 1990).
          }
\end{figure*}

We show in Figure~\ref{fig:Spec_PPN}
the aliphatic and aromatic C--H stretching features
of four PPNe. Except the Red Rectangle illuminated
by HD~44179 of $\Teff\approx7,750\K$,
PAHs in these PPNe are rich in aliphatic contents
and their aliphatic fractions are high,
with $\alifrac\approx25\%$ for IRAS~04296+3429,
$\alifrac\approx35\%$ for IRAS~05341+0852,
and $\alifrac\approx8.6\%$ for CRL~2688.
All these three aliphatic-rich PPNe are excited
by cool stars which lack UV photons
($\Teff\approx6,500\K$ for IRAS~04296+3429
and IRAS~03541+0852, and $\Teff\approx7,000\K$
for CRL~2688). This suggests that, in UV-poor regions,
once attained, PAHs are capable of maintaining
their aliphatic sidegroups without being stripped off.
Nevertheless, the Red Rectangle, also illuminated
by a cool star with little UV radiation,
shows very weak emission at 3.4$\mum$
and the PAH aliphatic fraction is only $\simali$0.3\%.
This indicates that, in addition to the ``hardness''
of the exciting stellar photons, some other factors
(e.g., the starlight intensity) are also at play
in affecting the PAH aliphatic fractions.

\begin{figure*}
\centering{
\includegraphics[scale=0.4,clip]{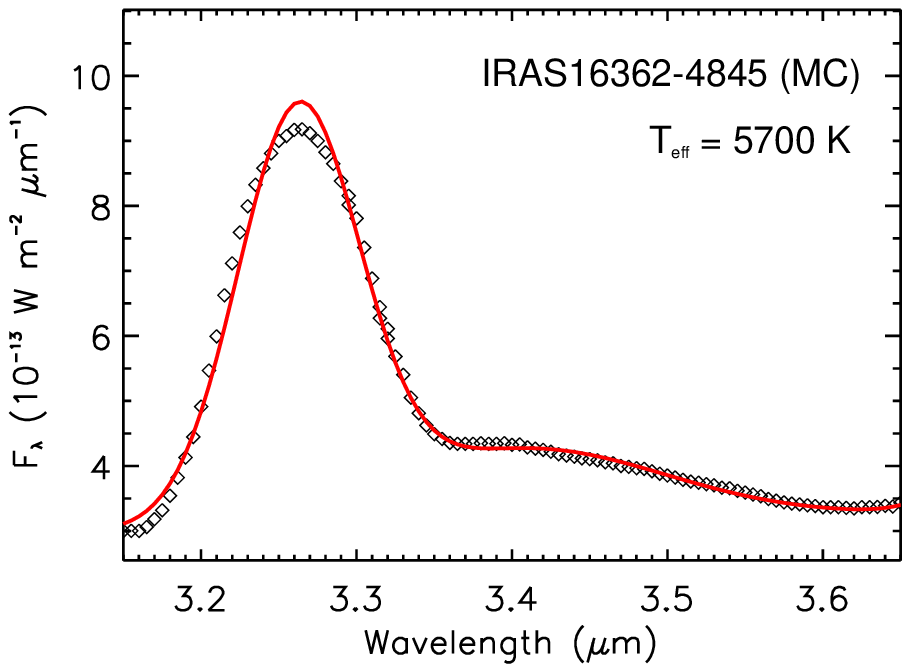}
\includegraphics[scale=0.4,clip]{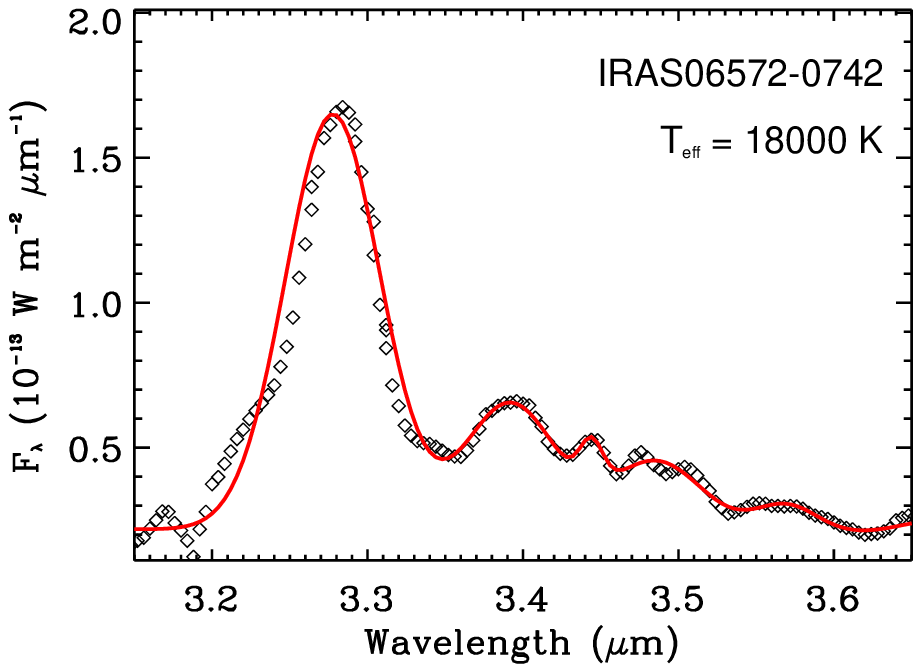}
\includegraphics[scale=0.4,clip]{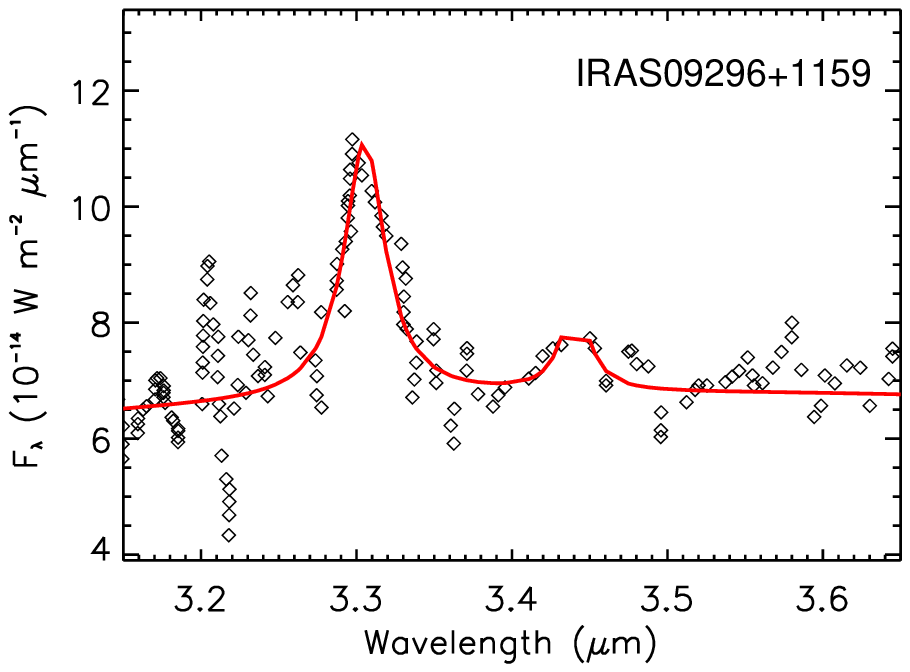}
\includegraphics[scale=0.4,clip]{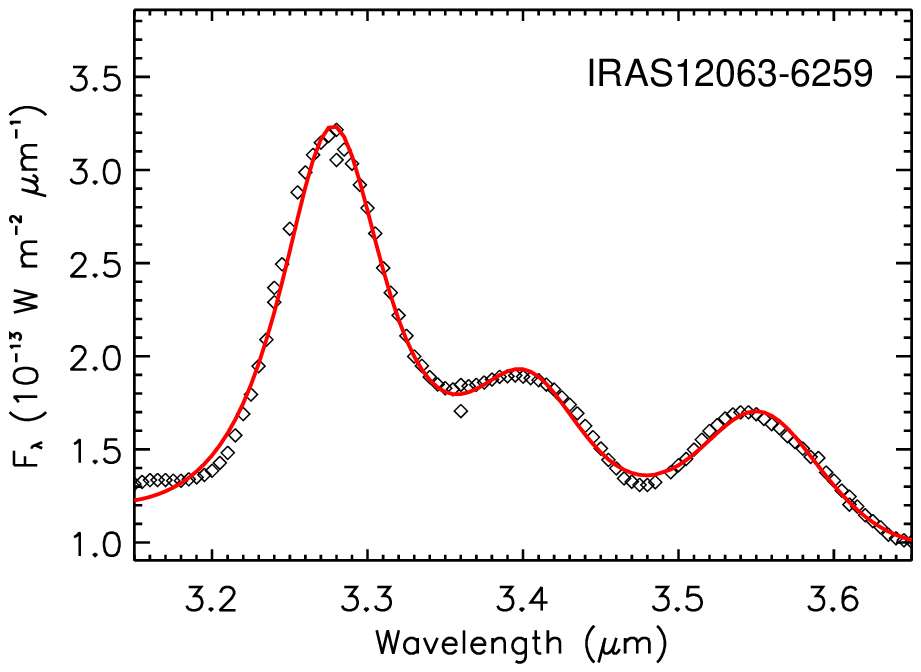}
\includegraphics[scale=0.4,clip]{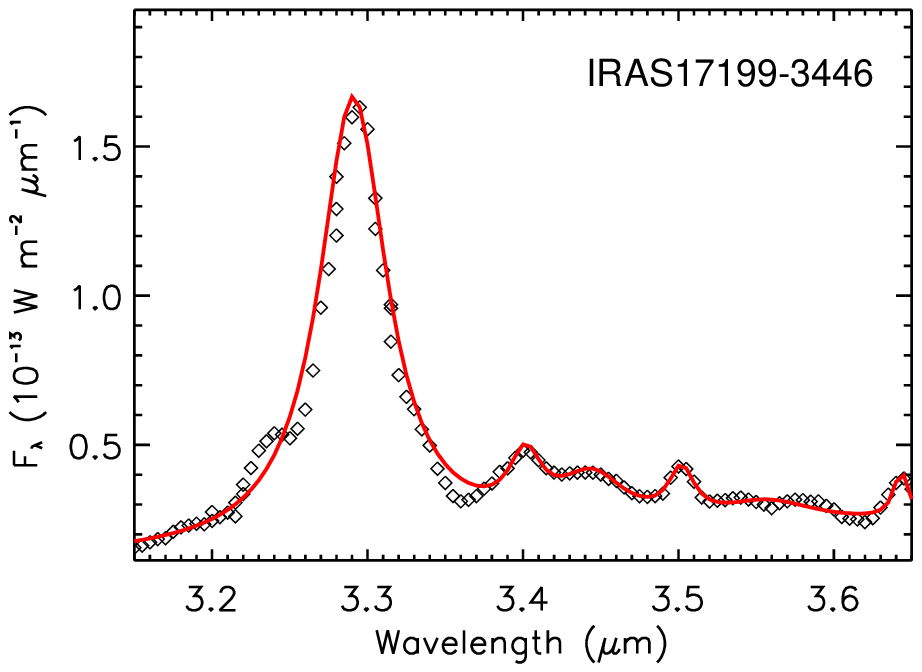}
\includegraphics[scale=0.4,clip]{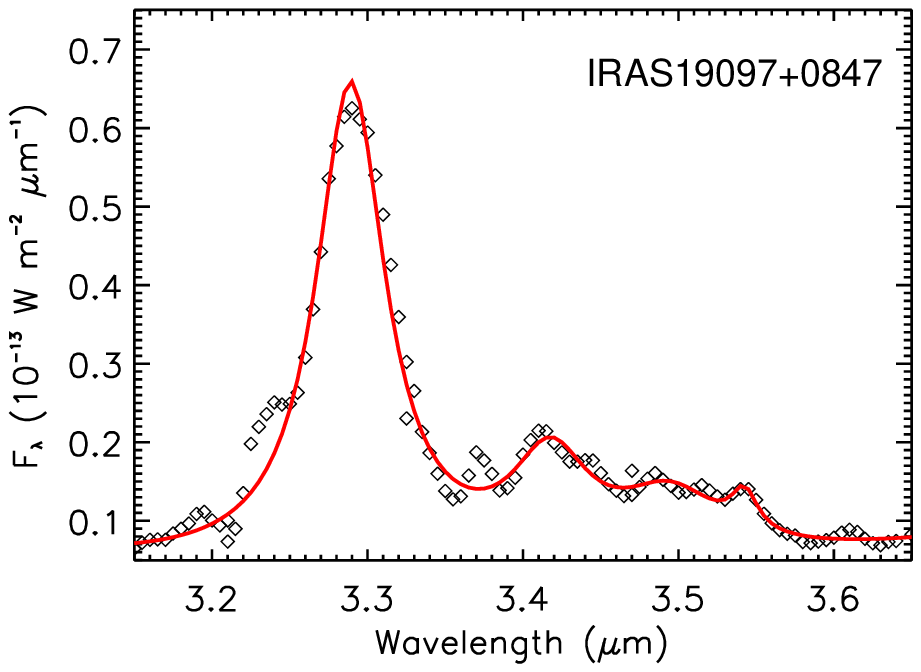}
}
\caption{\footnotesize
         \label{fig:Spec_unknown}
         Same as Figure~\ref{fig:Spec_PDR}
         but for six sources of which
         the effective temperatures of
         the illuminating stars are unknown:
         IRAS~16362-4845 (Jourdain de Muizon et al.\ 1990);
         IRAS~06572-0742 (Jourdain de Muizon et al.\ 1990);
         IRAS~09296+1159 (Geballe et al.\ 1990);
         IRAS~12063-6259 (Jourdain de Muizon et al.\ 1990);
         IRAS~17199-3446 (Jourdain de Muizon et al.\ 1990); and
         IRAS~19097+0847 (Jourdain de Muizon et al.\ 1990).
         }
\end{figure*}

Among our 24 Galactic sources, six objects lack
information about their illuminating stars.
Their near-IR spectra are shown
in Figure~\ref{fig:Spec_unknown}
and the intensity of the aliphatic
C--H stretching feature
(relative to the aromatic C--H feature)
varies substantially, from essentially
no 3.4$\mum$ emission in IRAS~16362-4845
to highly aliphatic in IRAS~12063-6259
($\alifrac\approx20.4\%$) and
in IRAS~06572-0742 ($\alifrac\approx6.0\%$).
IRAS~16362-4845 exhibits a smooth, flat
continuum at $\simali$3.4--3.6$\mum$.
The PAH aliphatic fractions
of IRAS~09296+1159 ($\alifrac\approx3.5\%$),
IRAS~17199-3446 ($\alifrac\approx4.8\%$),
and IRAS~19097+0847 ($\alifrac\approx4.6\%$)
are moderate.

\begin{figure*}
\centering{
\includegraphics[scale=0.4,clip]{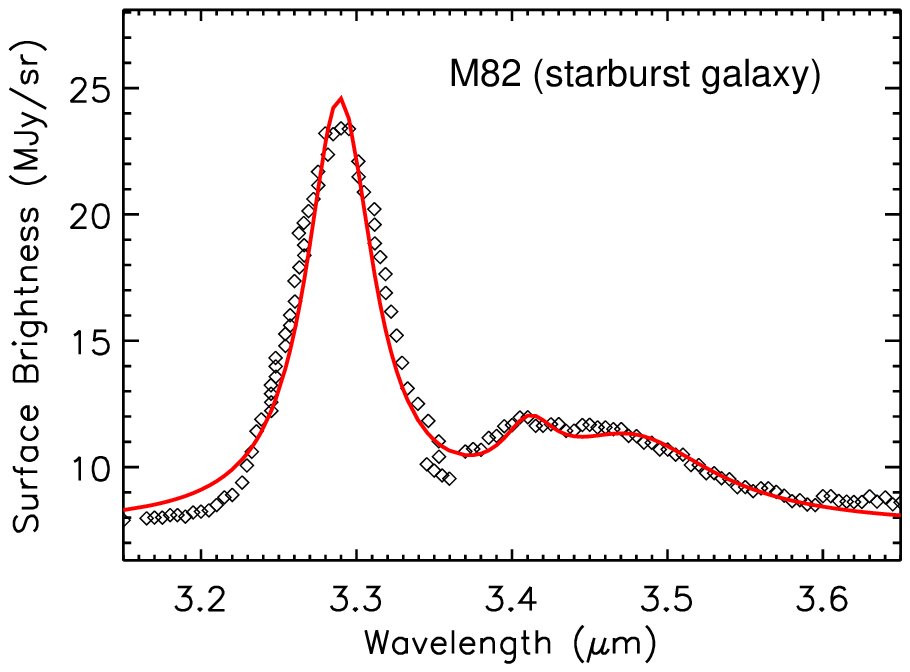}
\includegraphics[scale=0.4,clip]{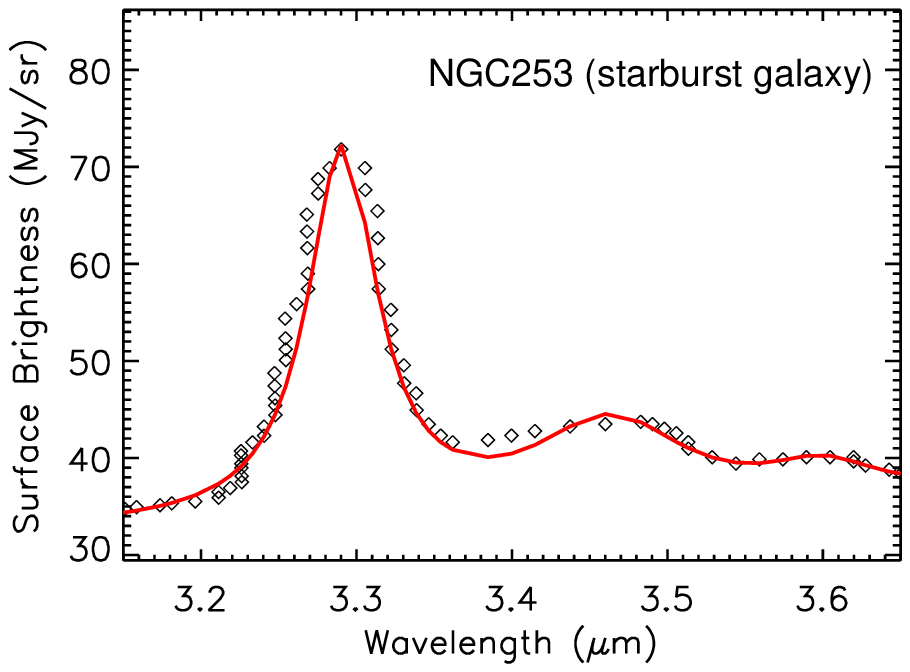}
\includegraphics[scale=0.4,clip]{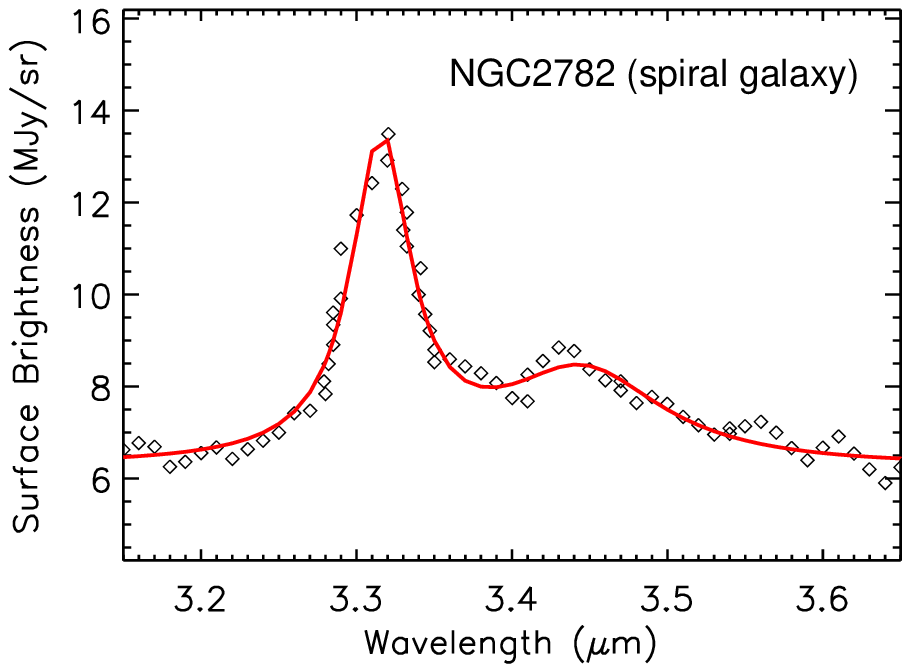}
\includegraphics[scale=0.4,clip]{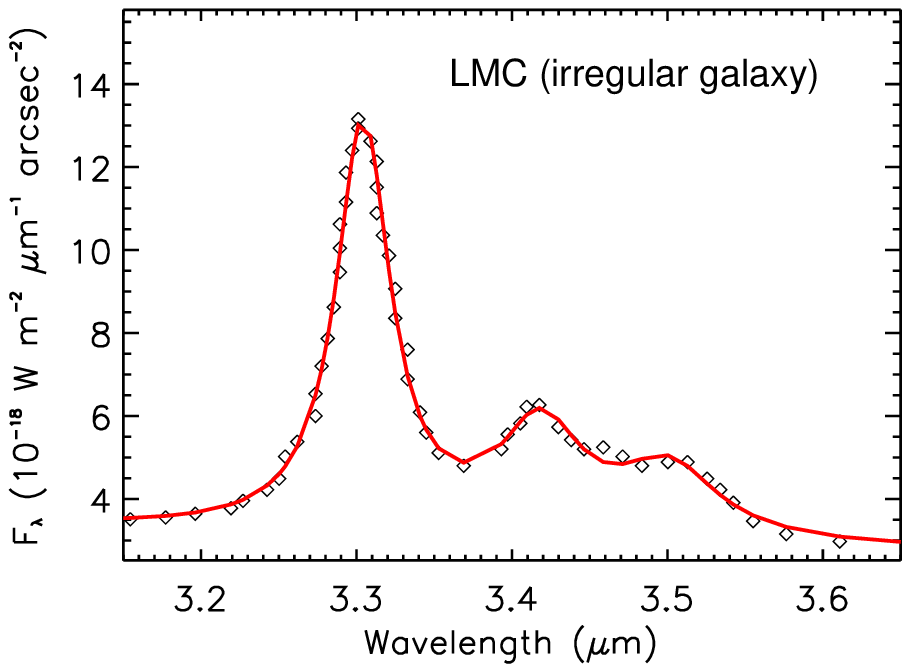}
}
\caption{\footnotesize
         \label{fig:Spec_galaxies}
         Same as Figure~\ref{fig:Spec_PDR}
         but for four external galaxies:
         M82 (Yamagishi et al.\ 2012);
         NGC~253 (Yamagishi et al.\ 2012);
         NGC~2782 (Onaka et al.\ 2018);
         and LMC (Mori et al.\ 2012).
          }
\end{figure*}

Figure~\ref{fig:Spec_galaxies} shows
the aliphatic and aromatic C--H stretches
of four nearby galaxies. They all show
considerable emission at 3.4$\mum$
and their PAH aliphatic fractions all exceed 7\%.
The Large Magellanic Cloud (LMC) and NGC~253,
a dusty starburst galaxy, also show a weak sub-feature
at $\simali$3.5 and 3.6$\mum$, respectively.
It is not clear if (and how) the galaxy metallicity
affects the PAH aliphatic fraction.
It has long been known that, since the ISO time,
the PAH abundance decreases as the metallicity
drops (see Li 2020 and references therein).
It is unclear if the presence and the intensity of
the 3.4$\mum$ feature (relative to the 3.3$\mum$
feature) are related to the metallicity.
In this respect, JWST, with its unprecedented
sensitivity and spatial resolution,
will allow an in-depth study.
In principle, in low-metallicity regions,
PAHs are less likely to attain aliphatic sidegroups
because of the low C/H ratio (and therefore low
--CH$_3$ abundance) and more likely to attain
extra H atoms to be superhydrogenated.
As the intrinsic strength of the 3.4$\mum$ aliphatic
C--H stretch of superhydrogenated PAHs is close
to that of PAHs with aliphatic sidegroups
(see Yang et al.\ 2020), observationally,
it is difficult to distinguish
whether the 3.4$\mum$ feature arises from
superhydrogenated PAHs or from PAHs with
aliphatic sidegroups.
Also, it is not clear if (and how) the 3.4$\mum$
feature is affected by the star formation rate.
We await  JWST for quantitative investigations.

\begin{figure}
\centerline
{
\includegraphics[width=12cm,angle=0]{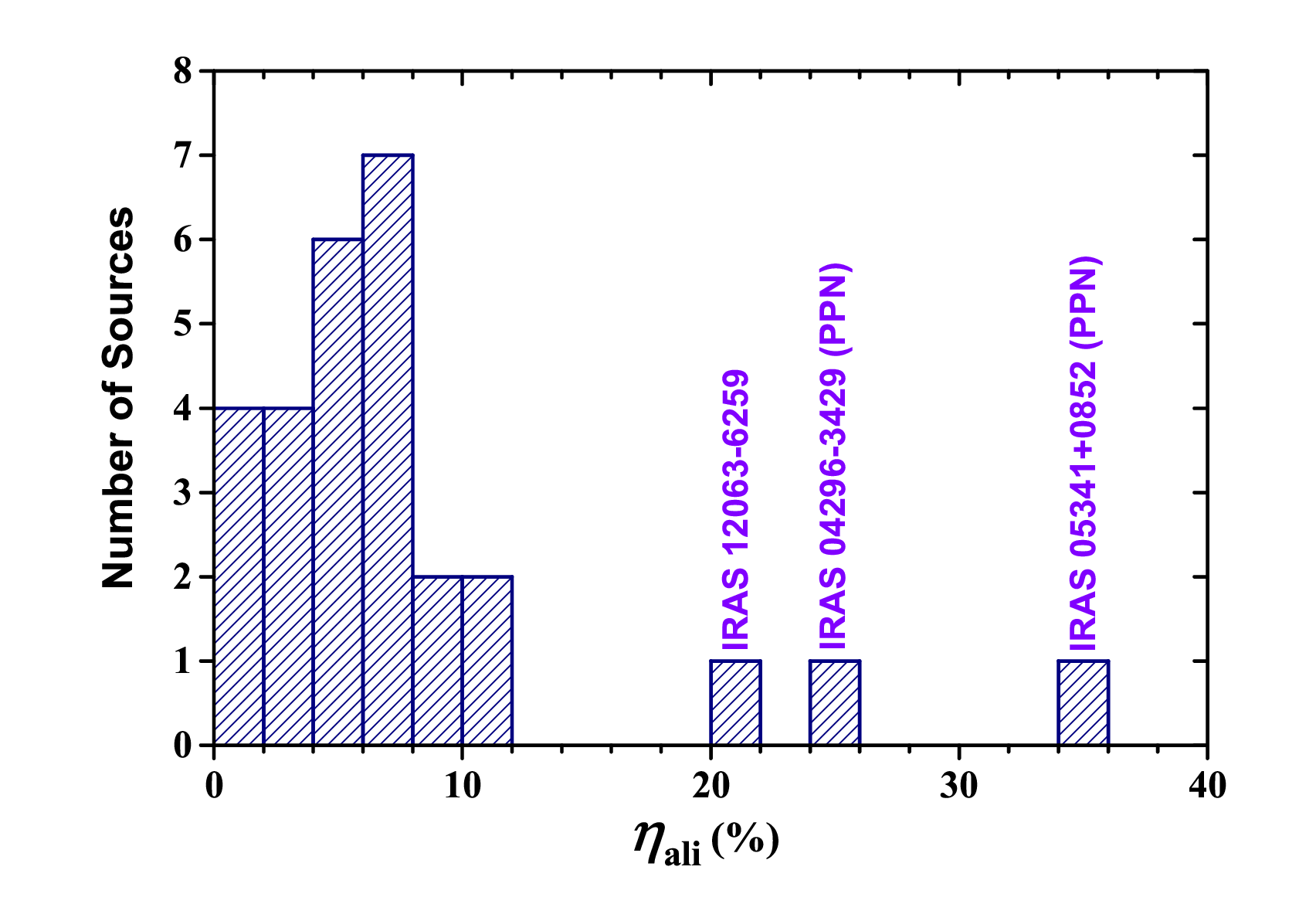}
}
\caption{\footnotesize
         \label{fig:histogram}
         Histogram of the aliphatic fractions
         of PAHs for 28 UIE sources.
         The median aliphatic fraction
         is $\langle \alifrac\rangle\approx5.4\%$.
         }
\end{figure}

In Figure~\ref{fig:histogram} we show
the PAH aliphatic fraction distribution
of our sample of 28 sources.
It can be seen that the majority (24/28) of these sources
has $\alifrac < 10\%$. The median of the PAH aliphatic
fraction  is $\langle \alifrac\rangle\approx 5.4\%$.
Two PPNe---IRAS\,04296+3429
($\alifrac\approx24.8\%$)
IRAS\,05341+0852
($\alifrac\approx34.9\%$)---are unusually
rich in aliphatics.
Another object---IRAS~12063-6259
($\alifrac\approx20.4\%$) of which
the exact nature is unknown---also
has a large aliphatic content.

\begin{figure}
 \vspace{-12mm}
  \begin{center}
\includegraphics[width=13.6cm,angle=0]{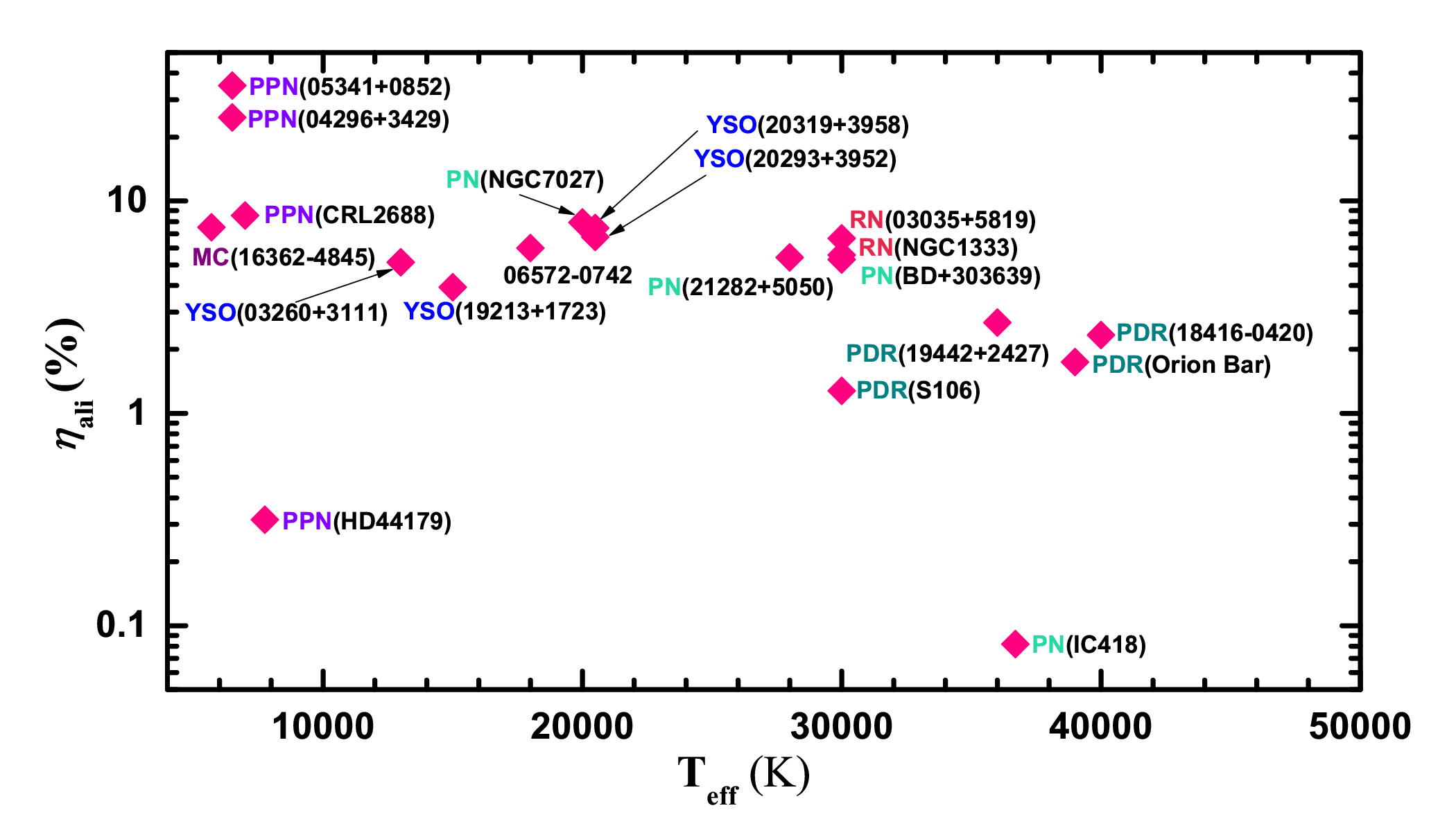}
  \end{center}
\vspace{-5mm}
\caption{\label{fig:Ali.vs.Teff} \footnotesize
               PAH aliphatic fraction ($\alifrac$)
               vs. stellar effective temperature ($\Teff$).
         PPN: protoplanetary nebula;
         PN: planetary nebula;
         RN: reflection nebula;
         MC: molecular cloud;
         PDR: photodissociated region;
         YSO: young stellar object.
	 }
\vspace{-3mm}
\end{figure}

We explore whether (and how)
the PAH aliphatic fraction varies
with astrophysical environments
(e.g., hardness of the exciting starlight photons).
As shown in Figure~\ref{fig:Ali.vs.Teff},
$\alifrac$ appears higher in regions
illuminated by stars with lower $\Teff$.
Indeed, as discussed above, it is generally
true that in UV-poor PPNe the 3.4$\mum$
emission feature (relative to the 3.3$\mum$ feature)
is much stronger than that of PNe, RNe, and PDRs
(see Figure~\ref{fig:Spec_PPN}).
Nevertheless, the Red Rectangle,
a PPN illuminated by HD~44179
of $\Teff\approx7,750\K$,
has a rather low $\alifrac$,
much lower than that of PDR and RNs
illuminated by stars of much higher $\Teff$.
This implies that, not only the hardness
but also the intensity of the starlight
may affect the accumulation and survival
of aliphatic sidegroups attached to PAHs.
This can be studied in more detail
by future JWST/NIRSpec observations
of spatially-resolved PAH spectra.
Previously, the spatial variations of $\Iratioobs$
with the exciting UV starlight intensities have
been investigated (e.g., see Joblin et al.\ 1996,
Sloan et al.\ 1997, Goto et al.\ 2003).
Again, with its unprecedented
sensitivity and spatial resolution,
JWST will allow us to explore the spatial
variations of the PAH aliphatic fractions
and their relations to the physical and chemical
conditions in an unprecedented depth.

While the 3.4$\mum$ aliphatic C--H emission is
widely seen in various astrophysical environments,
prior to JWST, the detection of the 6.85 and 7.25$\mum$
aliphatic C--H deformation bands is rare and has
so far been reported only in a couple dozen objects,
mostly based on the observations made with
the {\it Infrared Spectrograph} (IRS) on board
the {\it Spitzer Space Telescope} and
the {\it Shorter Wavelength Spectrometer} (SWS)
on board ISO (see Sloan et al.\ 2014, Yang et al.\ 2016a).
This will change with the launch of JWST:
due to its unprecedented sensitivity,
the MIRI spectrometer is well suited for
detecting the 6.85 and 7.25$\mum$ bands
(while the NIRSpec spectrograph is ideal for
detecting the 3.4$\mum$ band).
A combination of the 3.4, 6.85 and 7.25$\mum$ bands
would allow us to probe the aliphatic contents
of large PAHs (e.g., see Li \& Draine 2012).
It is interesting to note that,
in some planetary nebulae,
where the 3.4$\mum$ emission is observed,
an 5.25$\mum$ band is often also seen.
The 5.25$\mum$ band is thought
to be coming from large compact PAHs
(see Boersma et al.\ 2009).
Larger compact PAHs with aliphatic side groups
should be more eligible to survive in environments
illuminated by cool stars. It would be interesting to
consider, in the JWST era, large aliphatic PAHs
in a theoretical framework similar to
that presented in \S\ref{sec:model}
and see if any correlations exist
between these bands.

Finally, we note that the 3.4$\mum$ band could also arise
from superhydrogenated PAHs whose edges contain excess
H atoms (Bernstein et al.\ 1996, Sandford et al.\ 2013,
Yang et al.\ 2020). The addition of excess H atoms to PAHs
converts the flat sp$^2$ aromatic bonding of their asscoiated
C atoms into tetrahedral sp$^3$ aliphatic bonding,
resulting in the creation of aliphatic C--H stretching bands.
Compared with methylated PAHs
in which one aliphatic C atom
corresponds to three aliphatic C--H bonds,
for superhydrogenated PAHs, one ``superhydrogenated''
C atom corresponds to two aliphatic C--H bonds.
For superhydrogenated PAHs,
the ratio of the intensity of the 3.4$\mum$
aliphatic C--H stretch to that of the 3.3$\mum$
aromatic C--H stretch
$\langle\Aali/\Aaro\rangle\approx 1.98$
(Yang et al.\ 2020)
is similar to that of methylated PAHs
($\langle\Aali/\Aaro\rangle\approx 1.76$;
Yang et al.\ 2013, 2017b).
Therefore, the aliphatic fraction
as defined in eq.\ref{eq:alifrac3}
would be higher  by a factor of
$\approx \left(3/2\right)\times\left(1.76/1.98\right)
\approx1.33$, if the observed 3.4$\mum$ emission
is attributed to superhydrogenated PAHs.
%

%
%
%
%

\section{Summary}\label{sec:summary}
To facilitate a quantitative analysis of the aliphatic
and aromatic contents of PAHs in the JWST era,
we have proposed a theoretical framework
for determining the aliphatic fractions
($\alifrac$) of PAHs and have applied
the framework to pre-JWST UIE data.
Our major results are as follows:
\begin{enumerate}
\item An analytical formula for relating
          the PAH aliphatic fraction ($\alifrac$)
          to the emission intensity ratio of
          the 3.4$\mum$ feature
          to the 3.3$\mum$ feature ($\Iratio$)
          is presented. This relation is somewhat
          dependent on the ``hardness'' of the exciting
          stellar photons measured by the stellar
          effective temperature ($\Teff$).
\item To demonstrate the effectiveness of
          this framework (of deriving $\alifrac$
          from $\Iratio$), we have compiled
          the 3.3 and 3.4$\mum$ UIE data
          obtained in the pre-JWST era
          for an as complete as possible sample
          of 28 Galactic and extragalactic sources.
          We have then applied the $\alifrac$--$\Iratio$
          relation to these pre-JWST data.
\item We have derived the PAH aliphatic fraction
          $\alifrac$ for each source from the observed
          band ratio $\Iratioobs$ and found
          a median aliphatic fraction of
          $\langle\alifrac\rangle\approx 5.4\%$.
          Generally, the aliphatic fractions
          are the highest in protoplanetary nebulae
          illuminated by cool stars lacking UV radiation.
          However, the hardness of stellar photons
          is not the only factor affecting the PAH aliphaticity,
          other factors such as the starlight intensity
          may also play an important role.
\end{enumerate}

\acknowledgments{%
We thank B.T.~Draine, R.~Glaser,
A.N.~Witt and the anonymous referee
for valuable suggestions.
We thank J.S.~Spilker for providing us
the JWST data of SPT0418-47.
XJY is supported in part by
NSFC 12122302 and 11873041.
AL is supported in part by NASA grants
80NSSC19K0572 and 80NSSC19K0701.
}



\clearpage

\clearpage

\clearpage

\clearpage

\begin{table*}
\footnotesize
\begin{center}
\caption[]{\footnotesize
   Slopes for the Correlation between the Model
   Band Ratio $\Iratiomod$ versus $\NCali/\NCaro$
   for Neutrals and Cations
   as a Function of Stellar Effective Temperatures.
           }
\label{tab:Slope-all}
\begin{tabular}{lcc}
\noalign{\smallskip} \hline \hline \noalign{\smallskip}
$\Teff$ (K)	&	Neutral PAHs	&	Cationic PAHs	\\
\noalign{\smallskip} \hline \noalign{\smallskip}
3500	&	2.06	&	4.38	\\
6000	&	1.98	&	4.20	\\
10000	&	1.93	&	4.14	\\
22000	&	1.84	&	3.97	\\
30000	&	1.81	&	3.96	\\ \hline
Average	&	1.92	&	4.13	\\
Stdev	&	0.09	&	0.16	\\
\hline
\noalign{\smallskip} \noalign{\smallskip}
\end{tabular}
\end{center}
\end{table*}

\clearpage

\begin{table*}
\footnotesize
\begin{center}
\caption[]{\footnotesize
  Observed Band Ratios $\Iratioobs$ of Astronomical Sources
  Exhibiting Both the 3.3 and 3.4$\mum$ Emission, and
  PAH Aliphatic Fractions $\alifrac$ Derived from $\Iratioobs$
  Based on Eqs.\,\ref{eq:alifrac1}--\ref{eq:alifrac3}.
  }
\label{tab:Source-aliphacity}
\begin{tabular}{lccccc}
\noalign{\smallskip} \hline \hline \noalign{\smallskip}
Object	&	Type	&	$\Teff$	 &	Correlation	 &	$\Iratioobs$ 	 &	$\alifrac$ 	\\
        &           &    (K)     &   Slope               &             &   (\%)        \\ \hline
IRAS 18416-0420	&	PDR	&	40000	&	1.87 	&	0.15 	&	2.33 	\\
Orion Bar	&	PDR	&	39000	&	1.86 	&	0.11 	&	1.74 	\\
IRAS 19442+2427	&	PDR	&	36000	&	1.84 	&	0.17 	&	2.67 	\\
S106	&	PDR	&	30000	&	1.81 	&	0.08 	&	1.28 	\\
IC418	&	PN	&	36700	&	1.84 	&	0.01 	&	0.08 	\\
BD+303639	&	PN	&	30000	&	1.81 	&	0.34 	&	5.30 	\\
IRAS 21282+5050	&	PN	&	28000	&	1.81 	&	0.35 	&	5.42 	\\
NGC7027	&	PN	&	20000	&	1.84 	&	0.53 	&	7.94 	\\
IRAS 03035+5819	&	RN	&	30000	&	1.81 	&	0.43 	&	6.67 	\\
NGC 1333	&	RN	&	30000	&	1.81 	&	0.36 	&	5.58 	\\
IRAS 20293+3952	&	YSO	&	20500	&	1.83 	&	0.49 	&	7.47 	\\
IRAS 20319+3958	&	YSO	&	20500	&	1.83 	&	0.44 	&	6.78 	\\
IRAS 19213+1723	&	YSO	&	15000	&	1.87 	&	0.26 	&	3.93 	\\
IRAS 03260+3111	&	YSO	&	13000	&	1.90 	&	0.34 	&	5.15 	\\
HD44179	&	PPN	&	7750	&	1.97 	&	0.02 	&	0.32 	\\
CRL2688	&	PPN	&	7000	&	1.98 	&	0.62 	&	8.56 	\\
IRAS 04296+3429	&	PPN	&	6500	&	1.99 	&	2.18 	&	24.79 	\\
IRAS 05341+0852	&	PPN	&	6500	&	1.99 	&	3.56 	&	34.94 	\\
IRAS 16362-4845	&	MC	&	5700	&	2.00 	&	0.54 	&	7.54 	\\
IRAS 06572-0742	&	...	&	18000	&	1.85 	&	0.39 	&	6.00 	\\
IRAS 09296+1159	&	...	&	...	&	1.92 	&	0.24 	&	3.54 	\\
IRAS 12063-6259	&	...	&	...	&	1.92 	&	1.64 	&	20.37 	\\
IRAS 17199-3446	&	...	&	...	&	1.92 	&	0.32 	&	4.82 	\\
IRAS 19097+0847	&	...	&	...	&	1.92 	&	0.31 	&	4.59 	\\\hline
M82	&	starburst galaxy	&	...	&	1.92 	&	0.58 	&	8.35 	\\
NGC253	&	starburst galaxy	&	...	&	1.92 	&	0.52 	&	7.48 	\\
NGC2782	&	spiral galaxy	&	...	&	1.92 	&	0.81 	&	11.26 	\\
LMC	&	irregular galaxy	&	...	&	1.92 	&	0.72 	&	10.06 	\\
\hline
\noalign{\smallskip} \noalign{\smallskip}
\end{tabular}
\end{center}
\end{table*}
\clearpage

\end{document}